\documentclass[a4paper]{article}
\usepackage[utf8]{inputenc}
\usepackage{graphicx}
\usepackage{graphicx,wrapfig,lipsum}
\usepackage{array}
\usepackage[colorlinks=true,linkcolor=black, citecolor=blue]{hyperref}%
\usepackage{float, colortbl}
\usepackage{sans}
\usepackage{amsmath}
\usepackage{tabularx}
\usepackage{ragged2e}
\newcolumntype{L}{>{\RaggedRight\arraybackslash}X}
\newcolumntype{C}{>{\Centering\arraybackslash}X}
\newcolumntype{C}{>{\centering\arraybackslash}X}
\usepackage{booktabs}

\usepackage{xcolor}
\usepackage{dirtree}
\usepackage{newpxtext}

\usepackage{newpxmath}
\usepackage{bm}
\usepackage[caption=false]{subfig}
\usepackage[font={footnotesize},labelfont={bf}]{caption}
\usepackage{cite}
\usepackage{xcolor}

\newcommand{\beq}{\begin{equation}}
\newcommand{\eeq}{\end{equation}}
\newcommand{\bea}{\begin{eqnarray}}
\newcommand{\eea}{\end{eqnarray}}
\newcommand{\beqa}{\begin{eqnarray}}
\newcommand{\eeqa}{\end{eqnarray}}

\newlength{\dinwidth}
\newlength{\dinmargin}
\title{Design and optimisation of linear variable differential transformers and voice coil actuators using finite element analysis: a methodical approach to enhance sensor response and actuation force}
\author{
 Kumar. A. Kukkadapu$^a$, H. Van Haevermaet$^a$, Wim Beaumont$^a$, Nick van Remortel$^a$ \\ \\
$^a${\it  University of Antwerp, Particle Physics group,}\\ 
    {\it Groenenborgerlaan 171, 2020 Antwerpen, Belgium} \\ \\
}

\date{}

\begin{document}
\maketitle

\begin{abstract}
This study introduces a systematic and optimised methodology for designing Linear Variable Differential Transformer (LVDT) sensors and Voice Coil (VC) actuators, tailored for high-precision applications such as gravitational wave detectors and particle accelerators. Unlike prior studies, which focus primarily on industrial-grade LVDT design frameworks or isolated parameter studies, this work addresses the specific challenges of achieving both enhanced sensor response and actuation force within strict geometric and thermal constraints. Using a custom-developed simulation pipeline based on Finite Element Method Magnetics (FEMM), we evaluate the influence of key design parameters such as coil dimensions, radial gaps, and coil wire diameter on performance metrics such as response and linearity. The novelty of this work lies in its systematic exploration of design trade-offs, such as maximising performance while minimising heat dissipation, and its applicability to high-precision environments. In this work, particular emphasis is placed on the combination of the LVDT and VC functionalities in one unified sensor-and-actuator system designed for gravitational wave detectors. In addition, the methodology and simulation results are validated with experimental measurements of an optimised design, demonstrating a 2.8-fold increase in LVDT response and a 2.5-fold increase in VC actuation force compared to the initial configuration while preserving LVDT linearity and VC force stability. This work represents a significant advance over existing methodologies by offering a structured, scalable design process. 
\end{abstract}
\subsubsection*{keywords}{Linear Variable Differential Transformer (LVDT); Voice Coil (VC) actuator; finite element analysis (FEMM); sensor–actuator integration; gravitational wave detectors; seismic isolation systems; Einstein Telescope; ETpathfinder; high-precision instrumentation.} 
\section{Introduction}
\label{sec:introduction}

Linear Variable Differential Transformers (LVDTs) are highly sensitive and contactless linear position sensors extensively used in precision engineering applications. They are, e.g., widely used in industrial applications for position sensing in systems such as robotics, manufacturing, and~automotive controls \cite{commercialLVDT,Jefriyanto_2020}. Their working principle is based on mutual induction, and these commercial (or industry-grade) LVDTs consist of a movable ferromagnetic core sliding within a stationary assembly of one primary coil and two secondary coils. While exciting the primary coil with a high-frequency current, an axial displacement of the core induces a differential voltage in the secondary coils directly proportional to its position. This can also act as a transducer with self-actuating capabilities when DC currents are superposed on the high-frequency excitation signal \cite{commercialvc}. These transducers and actuators belong to a wider class of electromagnetic sensing and actuation techniques based on magnetic field coupling and inductive interactions. Similar principles are employed in various microwave, inductive, and~resonant sensing systems, where variations in geometry or material properties modulate the electromagnetic field distribution ~\cite{coplanar-hybradization-ex1, proximity-sensor-ex2, non-destructive-test-ex3, planar-microwave-ex4}. 

\begin{figure}[H]
\subfloat[\centering]{\includegraphics[width=0.47\linewidth]{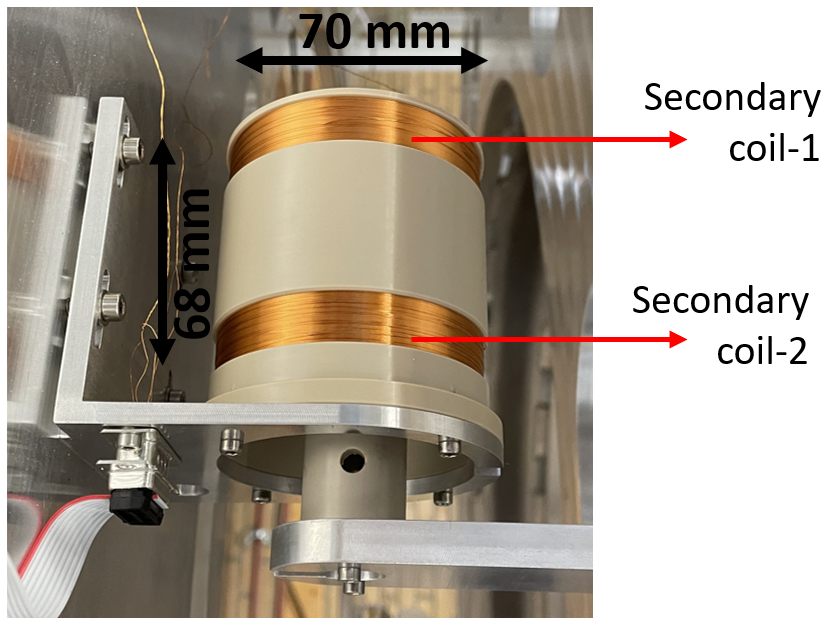}}
\subfloat[\centering]{\includegraphics[width=0.47\linewidth]{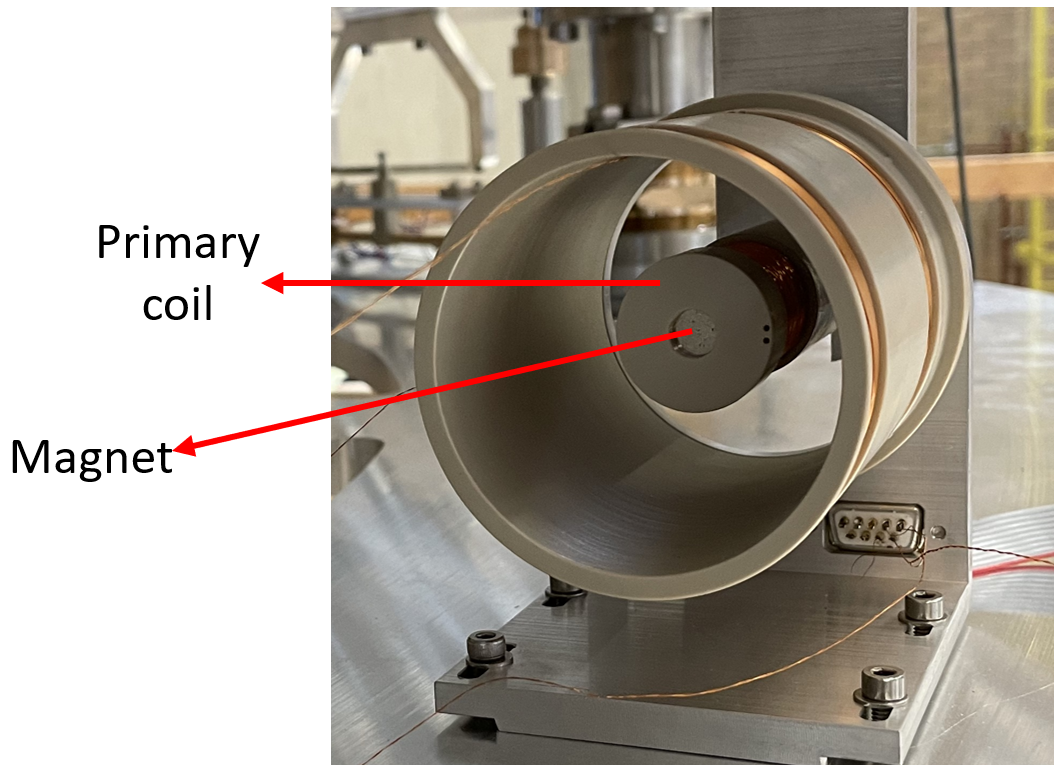}}
\caption{Example of an integrated LVDT+VC assembly installed in a seismic isolation stage of ETpathfinder~\cite{ETpathfinderTDR}. (\textbf{a}) Alignment of the moving primary coil relative to the fixed counter-wound secondary coils. (\textbf{b}) A view of the primary coil and magnet housing, showing the difference in primary and secondary coil radii to allow for residual transverse motion in the suspension. Both views illustrate the efficient design that enables simultaneous position sensing and~actuation.}
\label{fig:lvdt_vc}
\end{figure}

As commercial designs are optimised for robustness, operational range, and~cost effectiveness, they are generally not suitable for ground-based gravitational wave (GW) detectors that use position sensors and actuators in their seismic isolation \mbox{systems~\cite{STOCHINO2009737,Acernese_2014,Heijningen:2019jmd, 10.1063/1.4866659,Akutsu:2021auw,Kirchhoff:2020vhb,Punturo:2010zz}.} These require highly sensitive, low-noise, ultra-high vacuum (UHV) compatible devices. For~this purpose, special LVDT sensors~\cite{tariq2002linear} and Voice Coil (VC) actuators~\cite{WANG2002563} were developed, as~shown in Figure~\ref{fig:lvdt_vc}. A~key distinction between the LVDTs used in industry and GW detectors is the adoption of a moving primary coil instead of a moving ferromagnetic core. In~addition, the~radius of the primary is significantly smaller than the radius of the secondary coils to allow for a transverse motion of the suspended mass. The~movement of the excited primary (or inner) coil induces an electromotive force (emf) in the Maxwell pair arranged secondary (or outer) coils, which is detected as a differential voltage output. This contactless operation makes them ideal for environments demanding minimal mechanical interference, while the linear response ensures accurate position measurements. When a permanent magnet is placed inside the primary coil and the secondary coils are driven with a DC current, a~VC actuator functionality can be added. This creates a combined LVDT+VC system that enables position sensing and actuation with one device. The~key feature of the LVDT and VC are the linear response and the relatively stable actuation force, respectively.

\vspace{-3pt}

The method of varying individual design parameters to evaluate the effects on response, actuation, and~transfer characteristics has already been studied \cite{shar-model, sinshar-model, sharfadi, Mociran_Gliga_2023} before. However, the~methodology proposed in this paper offers a distinct contribution by introducing a structured optimisation method within a unified pipeline developed to simulate the performance of both LVDT and VC functionalities in a single framework, unlike the existing parametric studies that investigate individual design variables in isolation. We introduce a sequential approach to simultaneously optimise the design of LVDTs and VCs using finite element simulations. The~novelty of this work does not lie in the use of finite element simulations, but~in the systematic constraint-driven design process. This study provides a new procedure, starting from the design of the secondary coil and systematically moving towards the primary coil, to~obtain the best configuration within the existing mechanical~limits.

To demonstrate this methodical approach, a reference geometry was adopted from an existing LVDT+VC design used in ETpathfinder~\cite{ETpathfinderTDR, Utina:2022qqb}, whose dimensional values served as the basis for the simulations. ETpathfinder is an R\&D infrastructure for testing new GW detector technologies in a low-noise full interferometer environment, specifically for the Einstein Telescope (ET)~\cite{Punturo:2010zz}, a~future third-generation European GW detector. Due to the stringent sensing, actuation, and~noise requirements, ETpathfinder is the ideal test facility to illustrate how our methodology can improve the performance of LVDT+VC combinations. In~particular, using the new approach presented in this paper, we optimise the design of a specific LVDT+VC assembly that will be used in an inverted pendulum stage of the seismic isolation system. With~a dedicated experimental setup, a prototype is then fully characterised and validated with the simulation~results.   

This paper is structured as follows: Section~\ref{sec:simulation_framework} introduces the simulation framework that is used during the optimisation procedure, and~Section~\ref{sec:methodology} describes the proposed methodical approach. Section~\ref{sec:simulation_studies}  presents the simulation studies following this methodology, and~Section~\ref{sec:experimental_validation} shows the results of an experimental validation of the prototype developed for ETpathfinder, followed by Section~\ref{sec:conclusions} with the conclusions of our study. Throughout this paper, the~terms ‘primary coil’ and ‘inner coil’ are used interchangeably, as~are ‘secondary coil’ and ‘outer coil’.

\section{Simulation framework}
\label{sec:simulation_framework}

  The use of finite element methods to model LVDTs dates back three decades \cite{sykulski}. In~this study, we employ Finite Element Method Magnetics (FEMM)~\cite{femm}, an~open-source software package to conduct electromagnetic analysis in two-dimensional planar and axisymmetric domains. FEMM is capable of solving linear and non-linear magnetostatic, time-harmonic magnetic, electrostatic, and~steady-state heat flow problems. The~software features an interactive shell for defining problem geometry, material properties, and~boundary conditions with tools for meshing and solving the governing partial differential equations. For~the present study, simulations were executed using \texttt{pyFEMM 0.1.3} \cite{pyfemm}, a~\texttt{Python 3.10.1} extension to FEMM. Building on this foundation, we developed a dedicated simulation pipeline, which enables a systematic evaluation of multiple coupled parameters, including coil dimensions, radial gaps, wire diameter, and~thermal dissipation constraints, thereby establishing, for~the first time, a~completely structured optimisation methodology rather than isolated parameter~studies.

\begin{figure}[H]
\subfloat[\centering]{\includegraphics[width=0.33\linewidth]{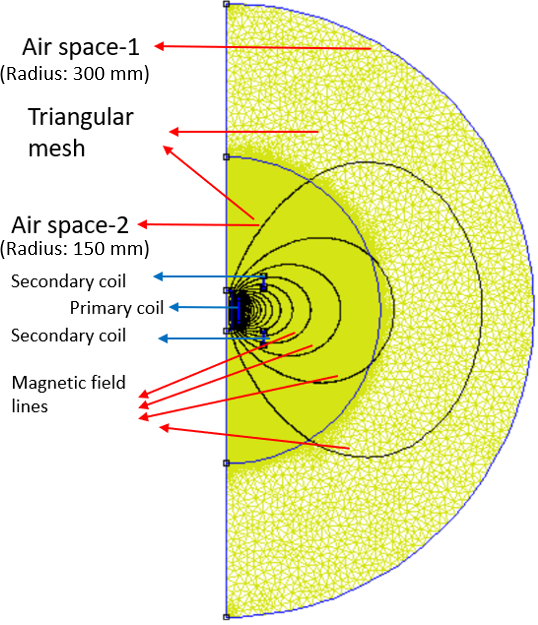}}
\qquad
\subfloat[\centering]{\includegraphics[width=0.26\linewidth]{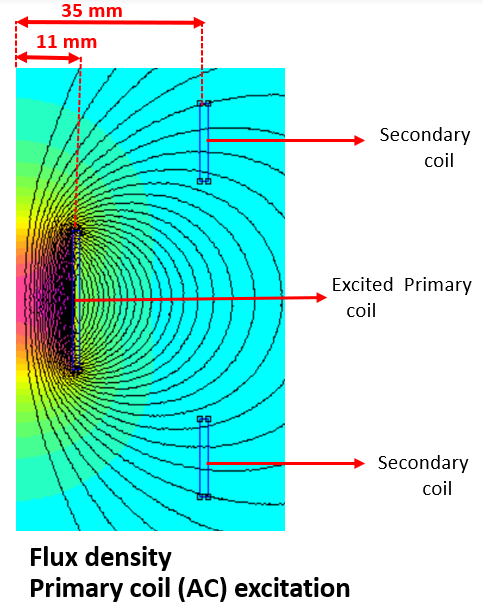}}
\qquad
\subfloat[\centering]{\includegraphics[width=0.31\linewidth]{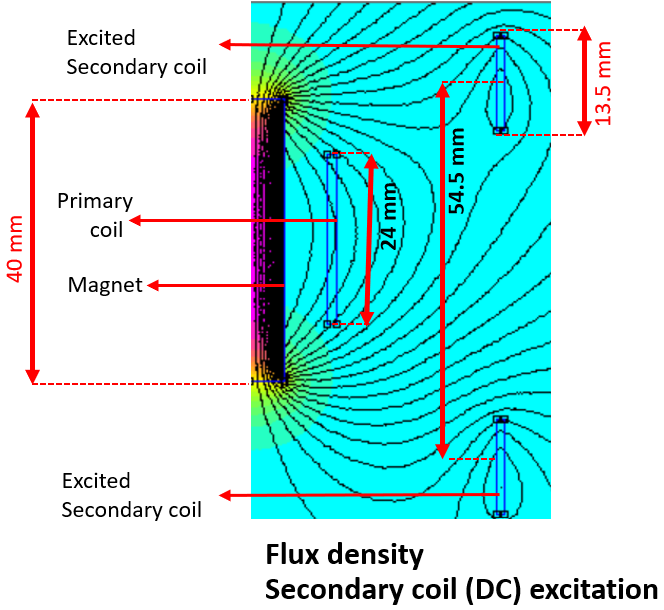}}
\caption{Axisymmetric FEMM simulation results for the integrated LVDT+VC system. (\textbf{a}) complete overview of the implemented model indicating the airspace volumes and different mesh sizes; (\textbf{b})~LVDT sensing mode with the primary coil excited at 10~kHz; (\textbf{c}) VC actuation mode with a DC current applied to the secondary coil, showing the resulting magnetic field distribution interacting with the permanent magnet to produce actuation force. Colour shading indicates magnetic flux density, while contour lines show the field lines linking primary and secondary coils. These results illustrate the flux paths in sensing versus actuation. The~indicated dimensions are defined in Table~\ref{tab:typeA_dimensions}.}
\label{fig:femm simulation}%
\end{figure}

The LVDT sensor is defined as a magnetostatic model in an axisymmetric $r$--$z$ plane, as~shown in Figure~\ref{fig:femm simulation}a, where $r$ represents the radial direction and $z$ the vertical direction. The~model plane is divided into two circular regions: Airspace-1 ($R_{\rm A1} = 300$~mm, triangular mesh, automatic size) and Airspace-2 ($R_{\rm A2} = 150$~mm, triangular mesh, size: 0.5) to model the ambient environment. Dirichlet boundary conditions are imposed at Airspace-1 with a null vector potential (to prevent the magnetic flux from entering the boundary). Inside Airspace-2, the LVDT coils and VC magnet are then positioned according to the design. For~all parts, the material properties were provided by the default FEMM library. 
For LVDT simulations, illustrated in Figure~\ref{fig:femm simulation}b, the~primary coil is excited with a 20~mA current at 10~kHz. This specific configuration is chosen based on the electronics used in ETpathfinder and the experimental setup (The
 amplification electronics are optimised for 10~kHz signals, while the Digital-to-Analogue Converter (DAC) used in the system provides a maximum output of 20~mA.). For~VC simulations, illustrated in Figure~\ref{fig:femm simulation}c, the~secondary coils are excited with a 1~A DC current. The~position of the primary coil is then varied relative to that of the secondary coils, and~the corresponding FEMM results are obtained at discrete intervals. To~obtain the LVDT response, complex voltages are extracted from the FEMM secondary coil circuit properties and used to calculate an absolute differential voltage signal. The~VC actuation is obtained from the axial component of the force computed via the steady-state weighted Maxwell stress tensor on the secondary coils~\cite{pyfemm}. This results in a series of data points that represent the output signal (LVDT response or VC force) at different positions. They can be used to produce an LVDT response profile (with the x-axis denoting the position of the primary coil relative to the secondary coil, and~the y-axis representing the differential voltage) and a VC force profile (with the y-axis representing the calculated force normalised to input current) to present the performance of the~system.

\vspace{-8pt}

\begin{table}[H] 
\caption{Design parameters of an ETpathfinder Type-A LVDT+VC combination~\cite{setup-paper}.}
\label{tab:typeA_dimensions}
\begin{tabularx}{\textwidth}{CC}
\toprule
\textbf{Parameter}	& \textbf{Value}\\
\midrule
Secondary coil distance ($\rm D_s$) & 54.5~mm\\
Secondary coil radius ($\rm R_s$) & 35.0~mm\\
Secondary coil height ($\rm H_s$) & 13.5~mm\\
Secondary coil layers ($\rm N_s$) & 7\\
Primary coil radius ($\rm R_p$) & 11.0~mm\\
Primary coil height ($\rm H_p$) & 24.0~mm\\
Primary coil layers ($\rm N_p$) & 6\\
Magnet radius ($\rm R_m$) & 5.0~mm\\
Magnet height ($\rm H_m$) & 40.0~mm\\
Magnet type & NdFeB~N40\\
Coil wire diameter ($\rm d_p, d_s$) & 32~AWG\\
\bottomrule
\end{tabularx}
\end{table}

\begin{figure}[H]
\subfloat[\centering]{\includegraphics[width=0.43\linewidth]{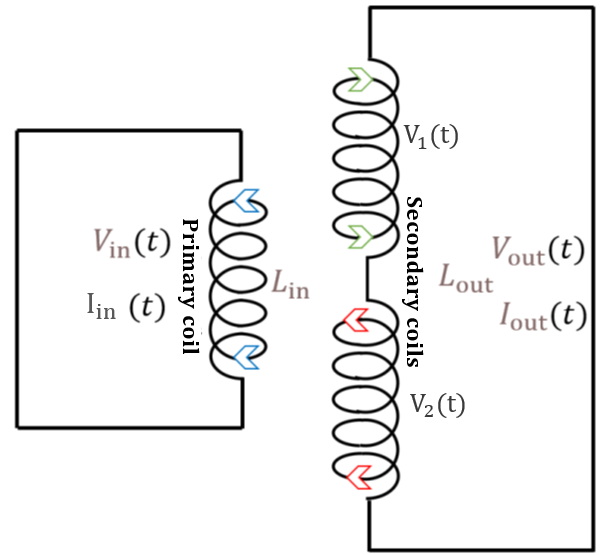}}
\qquad
\subfloat[\centering]{\includegraphics[width=0.52\linewidth]{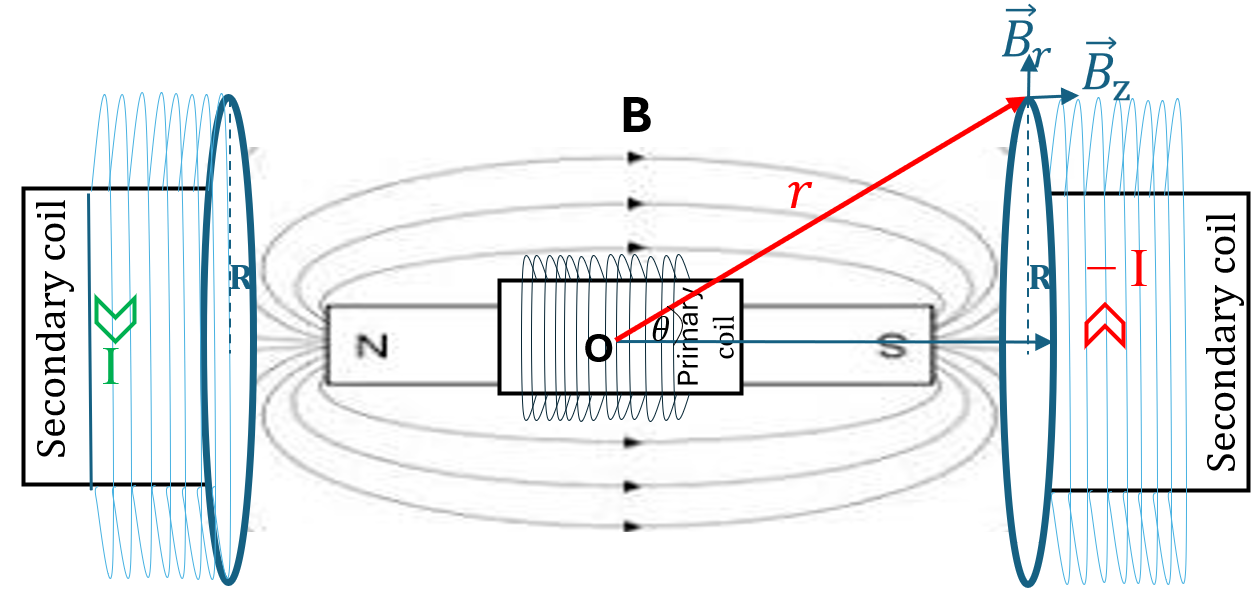}}
\caption{Equivalent electromagnetic representations of the LVDT and VC systems. (\textbf{a}) Transformer model of an LVDT with a primary coil and counter-winded secondary coils with inductances $\rm L_{in}$, $\rm L_{out}$, currents $\rm I_{in}$, $\rm I_{out}$ and potentials $\rm V_{in}$, $\rm V_{out}$. In this model, a coil displacement modifies the mutual inductance between primary and secondary coils. (\textbf{b}) Lorentz-force-based model of the VC actuator, where current-carrying coils ($\rm I$ and $\rm -I$) interact with the magnetic field $\rm \mathbf{B}$ of a permanent magnet to generate force. The length element $\rm d\mathbf{l}$ of one turn is located at distance $r$ with respect to origin O.}
\label{fig:lvdt and vc model}%
\end{figure}

To provide a clearer physical interpretation of the performance metrics, the~LVDT sensing and VC actuation mechanisms can be represented using equivalent electromagnetic models, shown in Figure~\ref{fig:lvdt and vc model}. In~the LVDT sensing mode, the~system can be approximated as a transformer with a primary excitation coil carrying an input current $\rm I_{in}$ and two secondary coils with voltages $\rm V_{1}$ and $\rm V_2$, where the induced differential voltage, $\rm V_{out} = |V_2 - V_1|$, depends on the mutual inductance variation with displacement. In~the VC actuation mode, the~system can be described by the Lorentz force generated by a current-carrying conductor in a magnetic field~\cite{Jackson:1998nia}. The~total force acting on a conductor can be expressed as: $\rm \mathbf{F} = I \int_L (\mathbf{B} \times d\mathbf{l})$, with $\rm I$ the current flowing through the conductor, $\rm \mathbf{B}$  the magnetic flux density vector generated by the permanent magnet, and~$\rm d\mathbf{l}$ the differential length vector along the total conductor path $\rm L$. For~the simplified case of a circular coil in a locally perpendicular uniform magnetic field, the axial force on a single turn of radius $\rm R$ can be described using the relation $\rm F_z= 2$$\pi$$ R I B_r$, where $\rm L=2$$\pi$$ R$ represents the length of the individual turn, $\rm B_r$ is the radial magnetic field component, and $\rm F_z$ the actuation force generated along the z-axis perpendicular to the coil. The~total force acting on both secondary coils is then obtained by summing the contributions from all turns, taking into account the spatial variation of the magnetic field and the orientation of each current~element.

{Based on these equivalent electromagnetic representations, the~following four key performance metrics are defined to quantitatively characterise the different LVDT+VC designs:} 

\begin{itemize}
    \item \textbf{LVDT Response 
 \Big($\rm \frac{V_{out}}{mm\cdot I_{in}}$\Big):} defined as the slope parameter of a first-order polynomial fit of the secondary coils differential voltage ($\rm V_{ out}$) as a function of the primary coil position. This is normalised with the input excitation current ($\rm I_{in}$) or voltage ($\rm V_{in}$). The~range of the fit function is chosen to avoid any bias from the non-linear LVDT response further away from the central point and~is in general smaller than the simulated or measured position range.  
    \item \textbf{LVDT Linearity (\%):} quantifies how closely the differential output voltage of the LVDT follows a linear function with the position of the moving coil and~is critical for applications requiring high precision. Any deviation from perfect 100\% linearity, also referred to as non-linearity, introduces a systematic error in the inferred position estimate. The~relation used to quantify the linearity $L({\rm z})$ is a relative fit residual:
\begin{equation}
    L({\rm z}) = \rm 100\times \frac{\lvert V_{fit}(z) - V_{data}(z) \rvert}{\lvert V_{data}(z) \rvert},
    \end{equation}
    where $\rm V_{data}(z)$ represents the simulated or measured differential voltage, $\rm V_{fit}(z)$ denotes the voltage calculated from the linear fit function, and~$\rm z$ is the position of the primary coil at which both are~obtained.

    \item \textbf{VC Force \Big($\rm \frac{F_z(z)}{I}$\Big):} the generated actuation force $\rm F_z(z)$ as a function of primary coil position $\rm z$, normalised to the DC input current I applied to the secondary coils. It is fitted with a second-order polynomial across the complete simulation range to extract the maximum force $\rm F_{max}$.
    \item \textbf{VC Force Stability (\%):} defined as a renormalised version of the VC Force ($\rm 100\times F_z(z)/F_{max}$), where~$\rm F_{max}$ is the maximum force obtained within the simulated or measured position range. It refers to the consistency and behaviour of the actuation force at different primary coil~positions.
\end{itemize}

These parameters, along with the resistance of the coils (that correlates directly with the heat dissipation), serve as the guiding metrics for the design and optimisation process. By~improving the LVDT response and linearity, and~simultaneously maximising the VC force and stability, the~overall system performance can be significantly~enhanced.

\section{Methodology}
\label{sec:methodology}

\begin{figure}[H]
    \centering
    \includegraphics[width=0.7\linewidth]{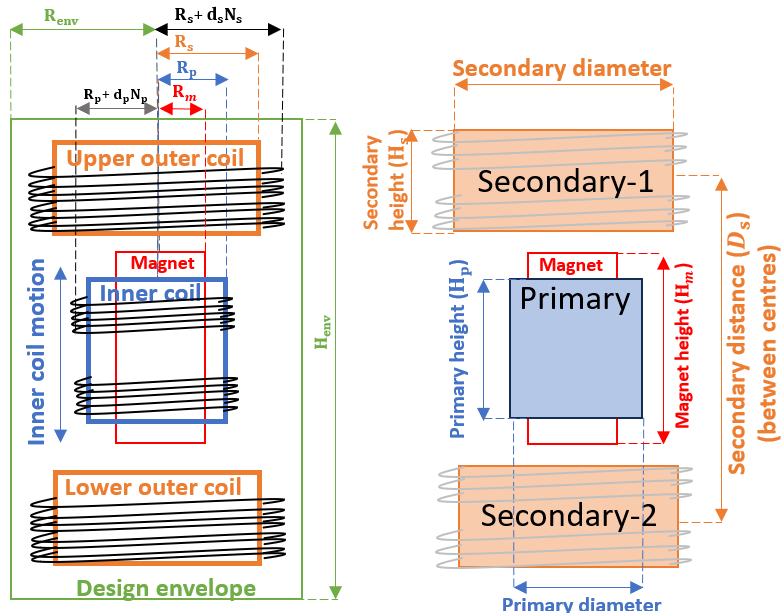}
    \caption{A drawing of a combined LVDT+VC system with all geometric design parameters~indicated.}
    \label{fig:lvdt geometry}
\end{figure}

The first step in the optimisation process is to establish the mechanical envelope imposed by the application requirements. The~complete LVDT+VC assembly must fit within a cylindrical volume of fixed outer radius ($\rm R_{env}$) and total height ($\rm H_{env}$), as defined in Figure~\ref{fig:lvdt geometry}. This boundary sets the maximum available space for the secondary coils. The~maximum size of the primary coil is then fixed by the required sensing and actuation range ($\rm \pm S$) along the coil axis and~a radial space ($ \Delta \rm R_{space}$) between the primary and secondary coils that ensures a sufficiently large transverse motion without mechanical interference between the inner (suspended) and outer~parts.

The design thus starts from a physically admissible initial configuration consistent with $\rm R_{env}, H_{env}, \pm S$ and $\Delta \rm R_{space}$. Subsequently, all coil parameters are iteratively refined based on the response metrics, defined in Section~\ref{sec:simulation_framework}, until~the target performance is met or~a constraint is hit. The~envelope defines what is possible, and~the response metrics determine what is optimal within that~boundary.

\vspace{-2pt}

Given a fixed outer envelope radius $\rm R_{env}$, the~radius of the secondary coil ($\rm R_{s}$) can be chosen close to this limit (after accounting for mechanical clearances), with~the following relation:
\begin{equation} \label{eq:start}
     \rm R_{s} + (d_{s}(N_{s}+ M_{s})) \leq R_{env},
\end{equation}
where $\rm d_{s} \ and \ N_{s}$ denote the diameter (including insulation) and number of layers of the secondary coil wire, respectively. In~addition, $\rm M_{s}$ is a safety margin scaled as an integral multiple of $\rm d_{s}$ to accommodate additional layers of wire beyond the nominal design, if~necessary. For~an ideal Maxwell pair configuration, the~distance $\rm D_s$ between the two secondary coils can be initialised using the following relation~\cite{tariq2002linear}:
\begin{equation} 
    \rm D_s = \rm \sqrt{3}R_{s}.
\end{equation}

This distance is measured between the two mid-points of each secondary coil, as~indicated in Figure~\ref{fig:lvdt geometry}. Based on these definitions, we can then initialise the height of the secondary coils ($\rm H_s$) using the following requirements:
\begin{equation}
    \rm H_s < D_s \quad and \quad D_s + H_s < H_{env}.
\end{equation}

When adjusting the height of a coil, we assume the convention that it is increased or decreased symmetrically with respect to its mid-point. Hence, for~larger $\rm H_s$ values, the actual separation between two coils will be smaller. Once the provisional secondary coil dimensions are determined, initial values for the primary coil and magnet dimensions can be chosen within the available space. With~a required sensing or actuation range $\rm \pm S$ (e.g., $\rm S=\pm 2.5$\,mm), a~conservative choice for the initial primary coil height ($\rm H_p$) is the smallest value satisfying:
\begin{equation} \label{eq:primary}
    \rm H_p \geq 2(S + M_s),
\end{equation}
where $\rm M_s$ represents an additional safety margin, chosen based on the use case (For tightly constrained motion, where the displacement is guaranteed not to exceed the sensing range, a~reduced margin of 5--10\% is sufficient. For~instance, in~a $\pm 5$~mm motion range, a~margin of $\pm 0.5$~mm is generally adequate.). Conversely, the~maximum allowed value is again determined by the envelope:
\begin{equation}
    \rm H_p < H_{env} - 2(S + M_s).
\end{equation}

This formulation ensures that the primary coil accommodates the full motion range with a buffer to account for mechanical tolerances and uncertainties. The~radius ($\rm R_p$) can be initialised with the following relation:
\begin{equation} \label{eq:end}
    {\rm R_s - R_{p} + d_{p}(N_{p}+ M_{p})} > \Delta \rm R_{space},
\end{equation}
where $\rm d_p \ and \ N_{p}$ are the diameter (with insulation) and number of layers of the primary coil wire respectively, and~$\rm M_{p}$ is an additional safety margin scaled as an integral multiple of $\rm d_p$ to accommodate additional layers of wire, if~necessary.
The radial gap $\Delta \rm R_{space}$ determines, together with $\rm R_s$, the~maximum primary coil radius and is defined by the user after accounting for all mechanical clearances. Note that additional radial space needs to be reserved for the material of the secondary coil bobbin, which can be, e.g.,~2~mm.

This baseline configuration serves as the starting point for the systematic optimisation method we introduce in this paper. It is a step-by-step procedure, starting from the design of the secondary coils and systematically moving towards the primary coil, magnet, and~finally the coil wire configuration. More specifically, we propose the following~sequence:
\begin{enumerate}
    \item Determine the optimal distance $\rm D_s$ between the secondary coils.
    \item Obtain the best secondary coil radius $\rm R_{s}$, and~subsequently the primary coil radius $\rm R_p$, based on the radial gap $\Delta \rm R_{space}$ requirements between the primary and secondary coils.
    \item Determine the optimal secondary coil height $\rm H_s$.
    \item Determine the optimal primary coil height $\rm H_p$.
    \item Maximise the magnet dimensions ($\rm R_m$ and $\rm H_m$) within the primary coil.
    \item Optimise the coil winding configuration: the coil wire diameter ($\rm d_p, d_s$) and~the number of layers ($\rm N_p, N_s$).
\end{enumerate}

This methodical approach ensures that all performance metrics of a combined LVDT+VC system are improved with respect to the initial design. It provides a clear and consistent instruction set that can be used to optimise any existing LVDT and/or VC configuration. By simultaneously increasing the performance of the sensing and actuation capabilities within the mechanical, electrical, and thermal boundaries, it is particularly suited for applications in GW detectors. The optimisation strategy adopted in this work follows a constraint-driven sequential approach rather than a global multi-objective optimisation scheme. This choice reflects practical design conditions, where geometric and thermal constraints impose a hierarchy on parameter~selection. 

To study the methodical optimisation in the following section and show the effects of changing various design parameters, a~reference geometry was adopted from an existing LVDT+VC design (Type-A) used in the seismic isolation system of ETpathfinder~\cite{setup-paper}. Starting from such a reference is a straightforward way to introduce the optimisation procedure, as~it anchors the parameter space to a proven configuration. In~the absence of a prior design, tentative dimensions for the primary and secondary coils can be obtained from Equations~(\ref{eq:start}) to \eqref{eq:end}. From~this reference geometry, or~an initial set of admissible dimensions, each of the parameters described above was tuned to examine its impact on the overall~performance.

\section{Simulation studies}
\label{sec:simulation_studies}
The simulations were performed with the software framework described in Section~\ref{sec:simulation_framework} and~the ETpathfinder Type-A LVDT+VC combination as defined in Table~\ref{tab:typeA_dimensions}. While alternative configurations have different absolute parameter values and quantitative variations, the~fundamental trends and underlying physical relationships are expected to remain unchanged. The~geometric parameters used throughout the analysis are defined according to the schematic shown in Figure~\ref{fig:lvdt geometry}.

\subsection{Secondary Coil~Design}
\begin{figure}[H]\vspace{-9pt}
\subfloat[\centering]{\includegraphics[width=0.49\linewidth]{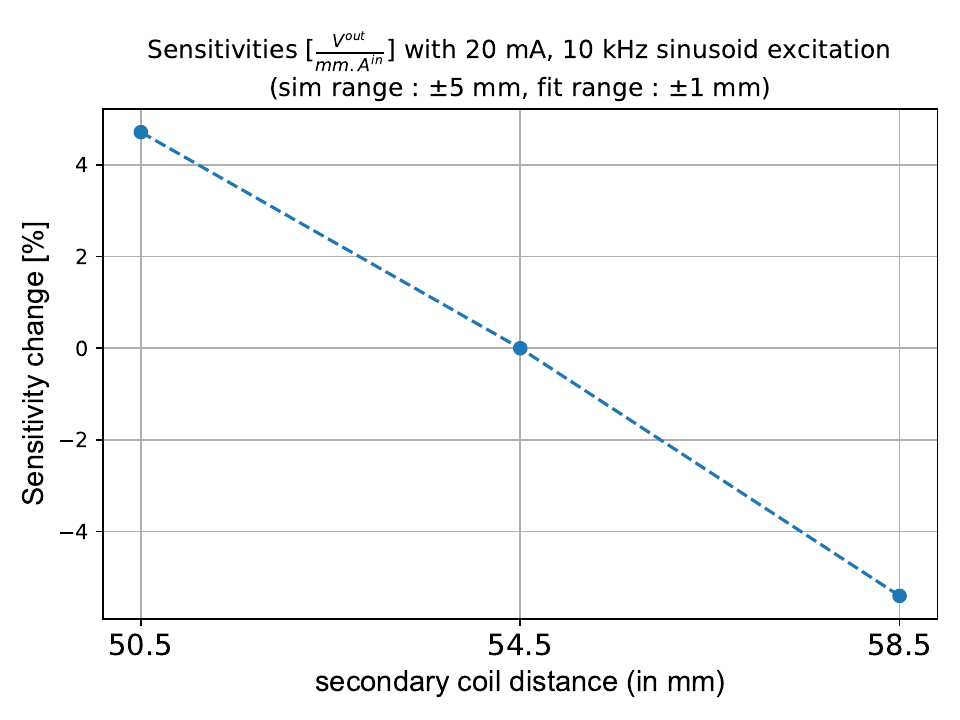}}
\subfloat[\centering]{\includegraphics[width=0.49\linewidth]{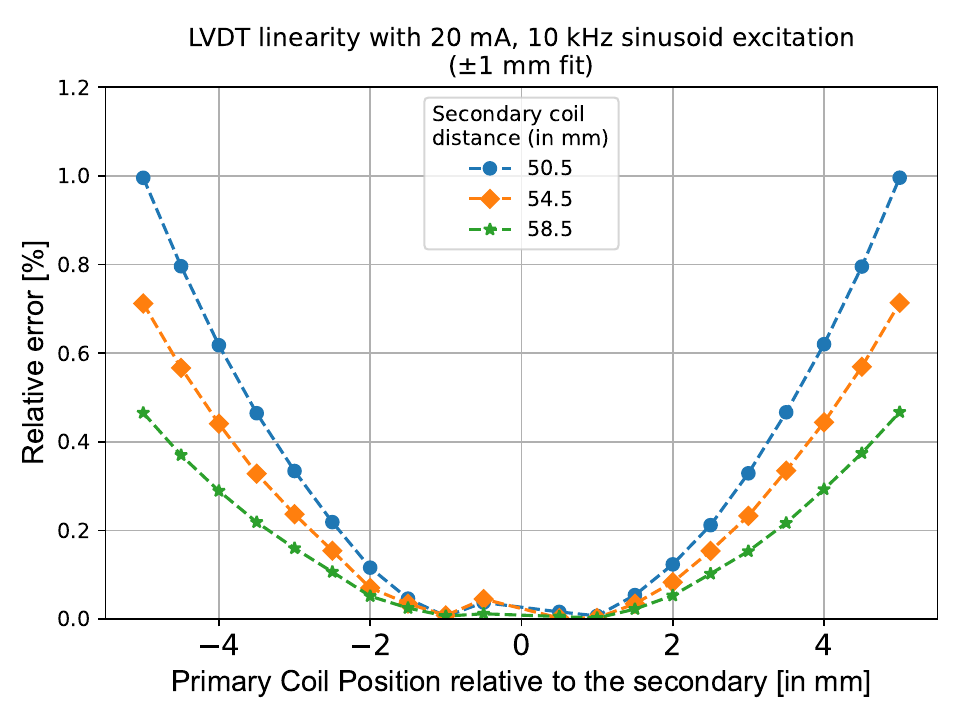}}
\caption{Effect of $\rm D_s$ on LVDT performance when the response is fitted within a $\pm 1$~mm range. (\textbf{a})~Response (V/mmA) increases slightly as $\rm D_s$ decreases, due to stronger mutual coupling. (\textbf{b})~Linearity degrades with reduced $\rm D_s$ beyond a $\pm 2$~mm displacement.}%
\label{fig:coil distance}%
\end{figure}

The optimisation process begins with the secondary coils. The~key parameters considered here are the distance between the two secondary coils ($\rm D_s$), the~radial gap relative to the primary coil ($\Delta \rm R_{space}$), and~the height of each secondary coil ($\rm H_s$). 
As shown in Figure~\ref{fig:coil distance}b, $\rm D_s$ directly impacts the linearity of the LVDT response, and~reducing the distance between secondary coils results in degraded linearity, especially after 2.5 mm of primary coil motion. However, it improves the response by 5\% per 4 mm reduction (Figure~\ref{fig:coil distance}a) in the chosen geometry. This is because closer coils couple more strongly to the flux of the primary coil, increasing the induced voltage. However, the~gain in response must be weighed against the increased non-linearity. If~the target application tolerates a non-linearity exceeding 1\%, closer coil spacing can be adopted. This provides greater flexibility in the overall design and lays the foundation for improving~response. 

\begin{figure}[H]\vspace{-14pt}
\subfloat[\centering]{\includegraphics[width=0.49\linewidth]{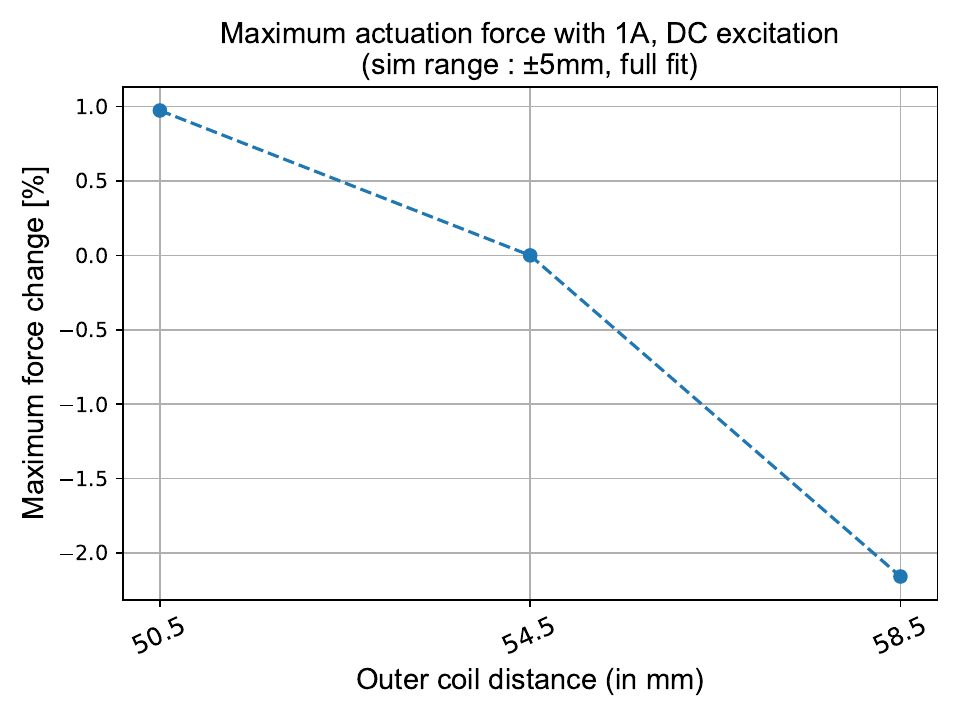}}
\subfloat[\centering]{\includegraphics[width=0.49\linewidth]{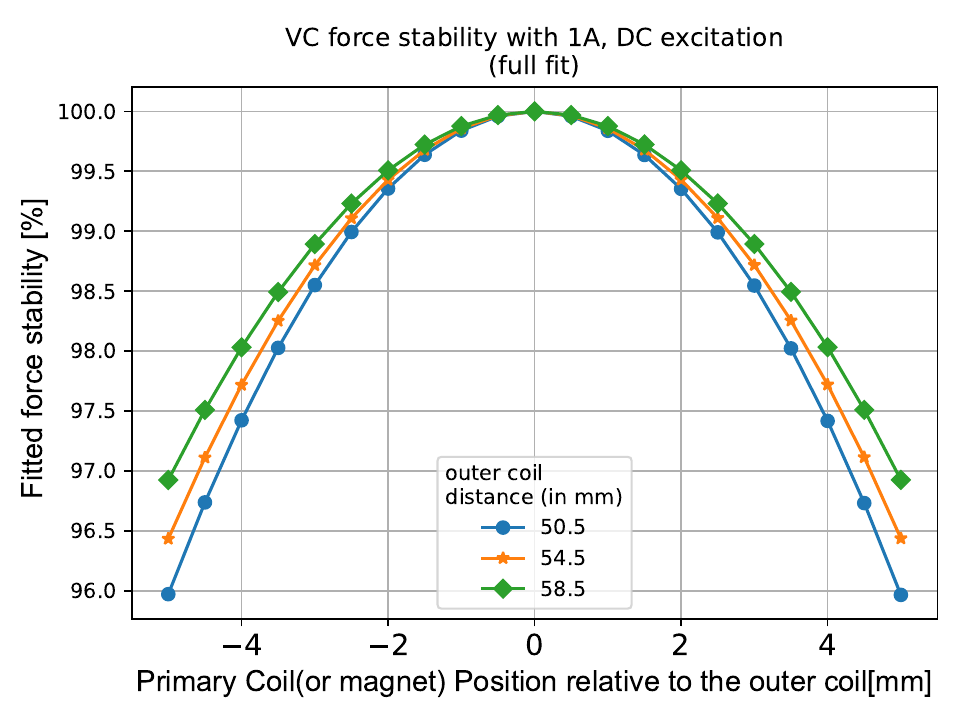}}
\caption{Influence of $\rm D_s$ on VC actuation force and stability with constant magnet dimensions. (\textbf{a})~Variation of $\rm F_{max}$ for different $\rm D_s$ values. (\textbf{b}) Corresponding change in VC force stability. Increasing $\rm D_s$ slightly reduces the maximum available force while marginally improving~stability.}
\label{fig:sec dist vc}
\end{figure}

The impact of $\rm D_s$ on the VC force is smaller than on the response, meaning that the distance can be optimised primarily for the LVDT functionality. As~shown in Figure~\ref{fig:sec dist vc}a, decreasing $\rm D_s$ by 4~mm increases the maximum actuation force by 1\%, while increasing $\rm D_s$ by the same amount reduces it by roughly 2\%. The~stability of the VC force, shown in Figure~\ref{fig:sec dist vc}b, decreases with a smaller value of $\rm D_s$: at $\pm 5$~mm, a 1\% loss is observed for a 8~mm decrease in secondary coil distance. The~custom-developed simulation pipeline can thus be used to identify the optimal $\rm D_s$ value that provides the best compromise between a high linearity and an enhanced response for the LVDT while preserving sufficient force stability for VC~operation. 

The second step in the design process is to tune the radial gap between the primary and secondary coils. Maintaining a minimal separation between them is essential for maximising the magnetic flux reception by the secondary coil. However, this proximity must be balanced with mechanical constraints on $\Delta \rm R_{space}$ to ensure adequate space for a transverse motion of the primary coil. The~gap can be reduced either by increasing $\rm R_p$ (bringing it closer to the secondary coil) or by decreasing $\rm R_s$ (bringing the secondary coils closer to the primary coil).

\begin{figure}[H]\vspace{-16pt}
\subfloat[\centering]{\includegraphics[width=0.49\linewidth]{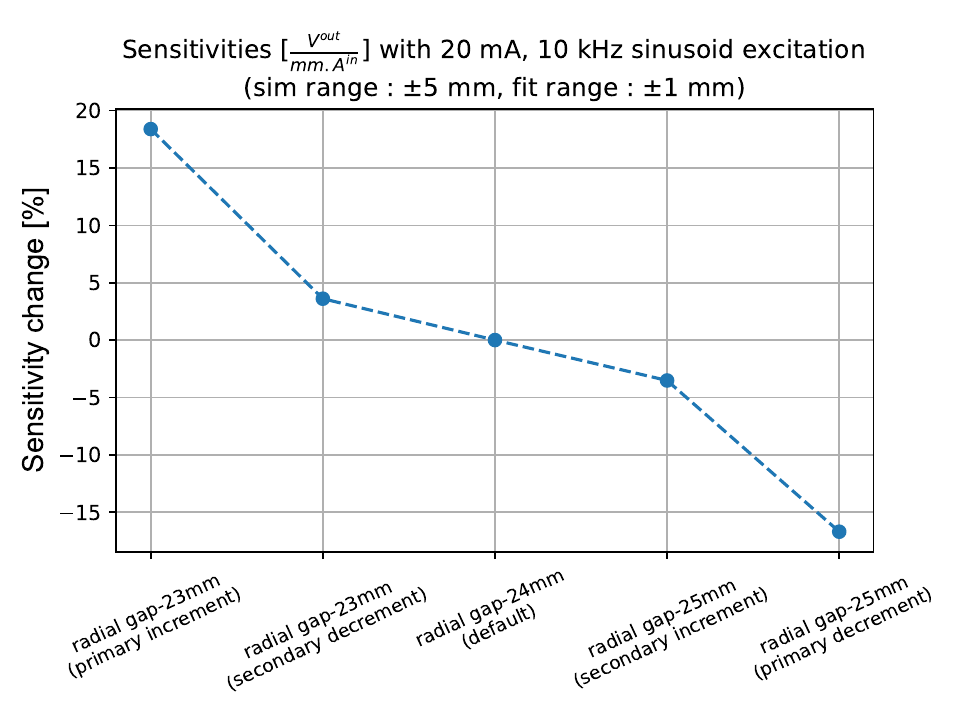}}
\subfloat[\centering]{\includegraphics[width=0.49\linewidth]{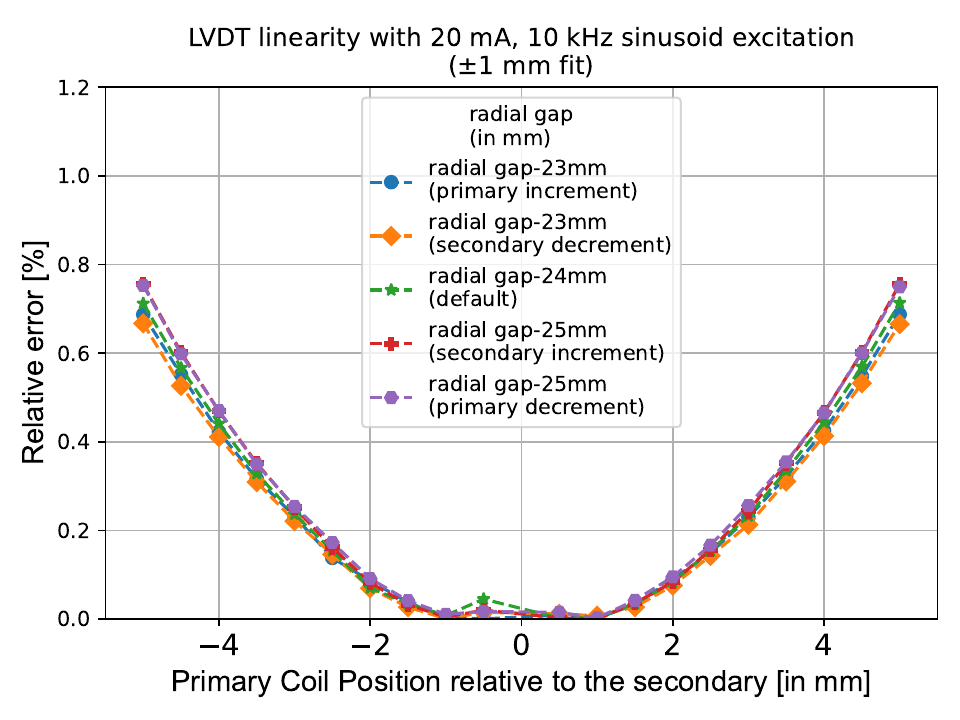}}
\caption{Effect of varying $\Delta \rm R_{space}$ on the LVDT performance when the response is fitted within a $\pm 1$~mm range for two cases: (1) increasing $\rm R_p$ and (2) reducing $\rm R_s$. (\textbf{a}) Response improves significantly when $\rm R_p$ increases and $\Delta \rm R_{space}$ is reduced. (\textbf{b}) Linearity remains largely unaffected for all~cases.}
\label{fig:radial gap}
\end{figure}

\begin{figure}[H]
\subfloat[\centering]{\includegraphics[width=0.49\linewidth]{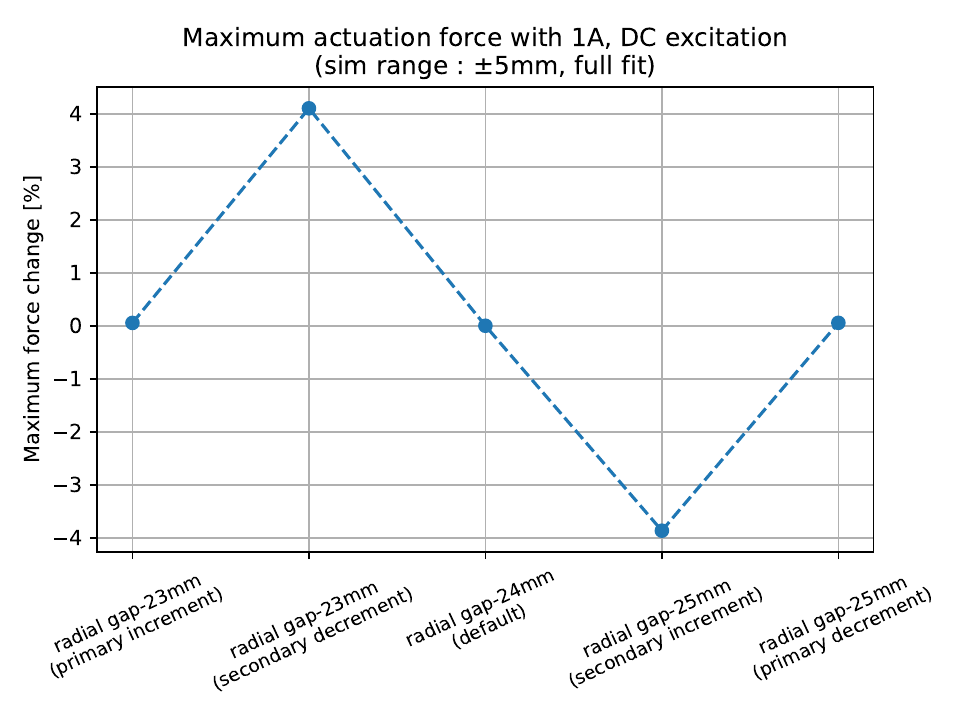}}
\subfloat[\centering]{\includegraphics[width=0.49\linewidth]{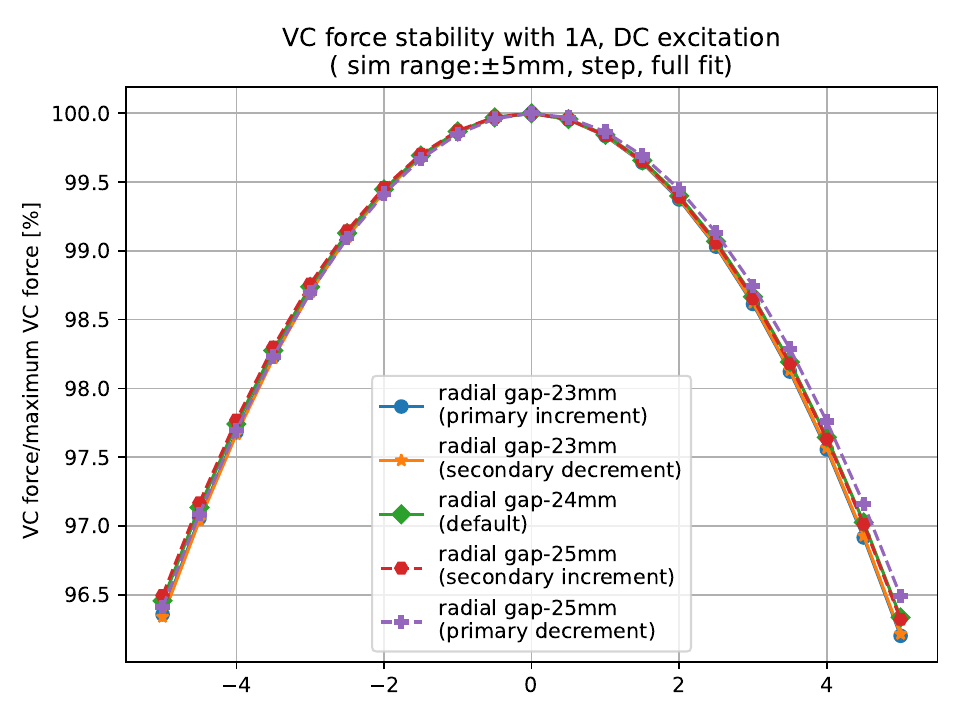}}
\caption{Influence of $\Delta \rm R_{space}$ on VC actuation force and stability with constant magnet dimensions. (\textbf{a}) reducing $\rm R_s$ increases the maximum VC force by~enabling a stronger interaction with the magnetic field of the permanent magnet. Changing $\rm R_p$ has no effect. (\textbf{b}) VC force stability remains essentially unchanged in both cases, indicating that an optimisation of $\Delta \rm R_{space}$ can improve both the LVDT or VC performance without compromising~stability.}
\label{fig:vc_force_radialgap}
\end{figure}

While increasing the primary coil radius enhances the LVDT response by 18\% per mm, as shown in Figure~\ref{fig:radial gap}a, it also significantly raises the resistance and thus power dissipation. Therefore, the~thermal requirements of the system must be considered. Conversely, reducing the secondary coil radius provides a safer alternative, as~it avoids increased heat generation while achieving a moderate improvement in response of 4\% per mm. Considering the VC actuation function, and~assuming constant magnet dimensions, we see that reducing the secondary coil radius also improves the actuation force, whereas changing the primary coil radius does not affect the force as one would expect. As~shown in Figure~\ref{fig:vc_force_radialgap}a, a~1 mm reduction in $\rm R_s$ increases $\rm F_{max}$ by approximately 4\%. No significant change in LVDT linearity (Figure \ref{fig:radial gap}b) and VC force stability (Figure \ref{fig:vc_force_radialgap}b) is observed when varying $\Delta \rm R_{space}$ within 2~mm. 

\vspace{-19pt}

This presents a trade-off: prioritise a significant improvement in LVDT response (by increasing the primary coil radius) at the cost of thermal penalty, or~achieve a balanced improvement in both LVDT and VC performance (by reducing the secondary radius). The~choice depends on the requirements of the~application.

\begin{figure}[H]\vspace{-15pt}
\subfloat[\centering]{\includegraphics[width=0.49\linewidth]{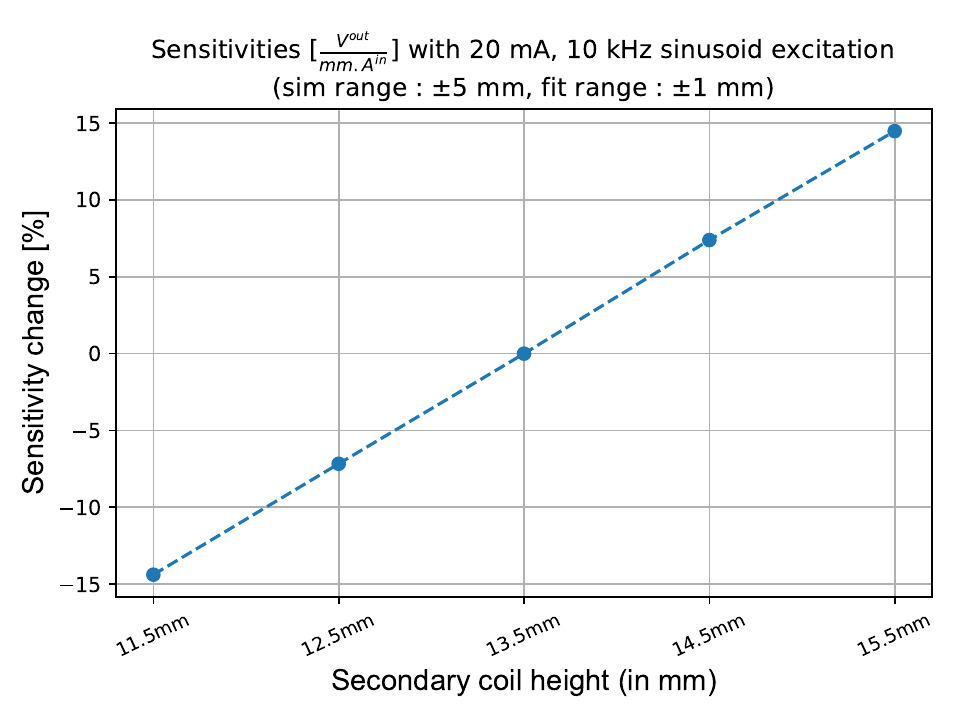}}
\subfloat[\centering]{\includegraphics[width=0.49\linewidth]{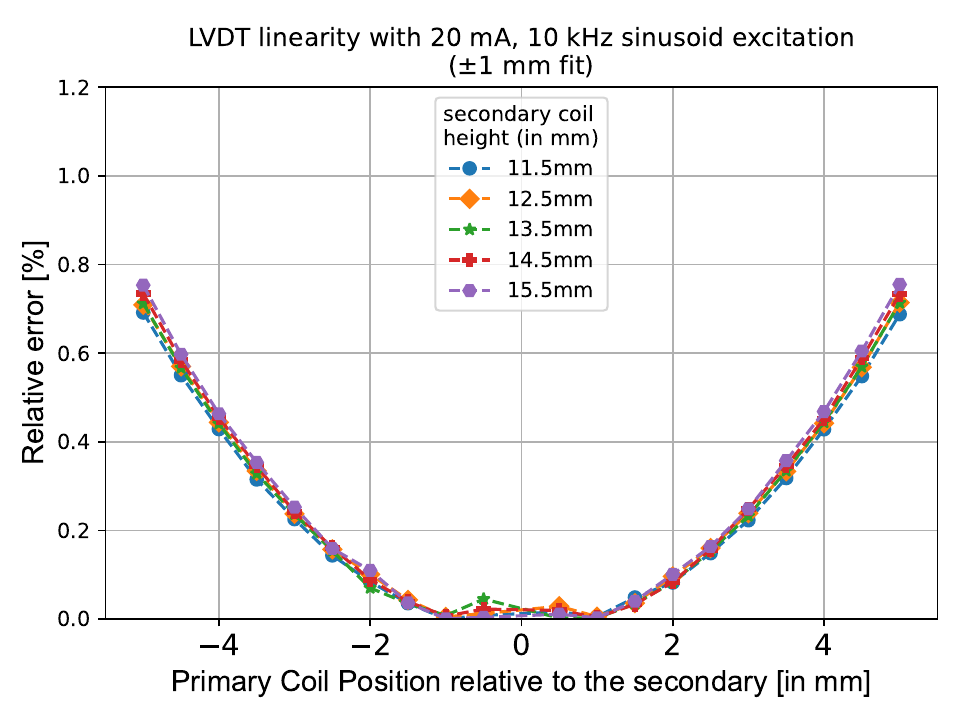}}
\caption{Effect of $\rm H_s$ on the LVDT performance when the response is fitted within a $\pm 1$~mm range. (\textbf{a}) Response increases approximately linearly with $\rm H_s$ due to a larger flux linkage. (\textbf{b})~Linearity remains unaffected, allowing $\rm H_s$ to be maximised within spatial limits (i.e., $\rm H_{env}$) without compromising~accuracy.}
\label{fig:lvdt response outer ht}
\end{figure}

After establishing the secondary coil distance and the radial gap in the initial phase, the~next step focuses on determining the optimal height of the secondary coils ($\rm H_s$).  The~LVDT response, directly related to magnetic flux linkage, is proportional to the height of the secondary coils, as demonstrated in Figure~\ref{fig:lvdt response outer ht}a. The~response scales approximately linearly with $\rm H_s$, improving by around 7.5\% per mm increase, while the LVDT linearity remains effectively constant (Figure~\ref{fig:lvdt response outer ht}b). By~maximising the height of the secondary coils within the envelope $\rm H_{env}$, the~design achieves the highest possible
LVDT response without compromising structural or spatial~limitations. 

\begin{figure}[H]\vspace{-15pt}
\subfloat[\centering]{\includegraphics[width=0.49\linewidth]{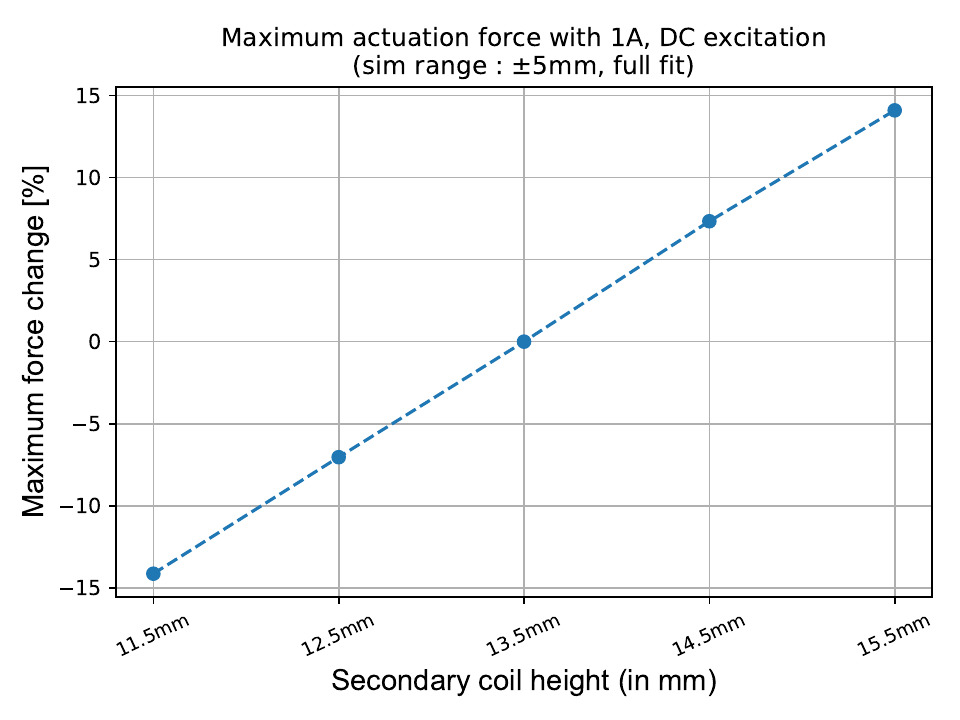}}
\subfloat[\centering]{\includegraphics[width=0.49\linewidth]{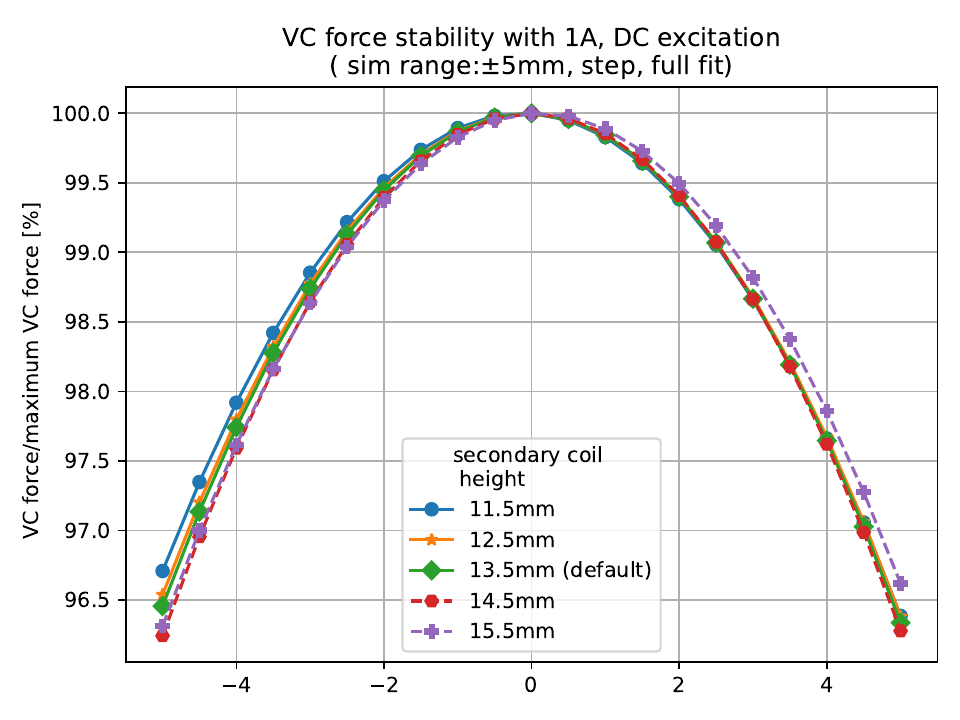}}
\caption{Effect of $\rm H_s$ on VC actuation force and stability with constant magnet dimensions. (\textbf{a})~Increasing $\rm H_s$ strengthens the actuation force by increasing the number of flowing charge carriers in the magnetic field. (\textbf{b}) VC force stability remains essentially unchanged, indicating that the $\rm H_s$ optimisation significantly benefits both LVDT and VC performance without drawbacks on linearity and~stability.}
\label{fig:vc response outer ht}
\end{figure}

Figure~\ref{fig:vc response outer ht} presents the corresponding relationship between the secondary coil height and VC actuation force. Increasing $\rm H_s$ also increases the maximum actuation force with an improvement of 7.5\% per mm (Figure~\ref{fig:vc response outer ht}a), which is essentially the same as the LVDT response improvement. In~addition, almost no impact on the VC force stability is visible (Figure~\ref{fig:vc response outer ht}b). This is because a wider coil allows more magnetic field lines from the permanent magnet to cut through the coil, allowing more current-carrying wire to interact with the magnetic field and~thereby boosting the actuation (or Lorentz) force. Thus, the~secondary coil height is typically maximised within mechanical~limits. 

\subsection{Primary Coil~Design} 
\label{sec:primary coil design}
The primary coil plays a pivotal role in determining the response of the LVDT. Its height and radius directly govern the magnetic flux generation that drives the secondary coil response. The~axial component of the magnetic field, $\rm B(t,z)$, at~the axis (i.e., $r = 0$) produced by the primary coil is obtained from the Biot--Savart law as follows~\cite{Jackson:1998nia}:
\begin{equation} \label{eq:mag-field-gen}
{\rm B(t,z)} = {\rm n_p} \frac{\mu_0 {\rm I_p(t) R_p^2}}{\rm 2\left(z^2+R_p^2\right)^{3 / 2}} {\rm \mathbf{e}_z},
\end{equation}
where $\rm z$ is the axial distance from the primary coil centre, $\rm n_p$ is the number of turns, $\rm R_p$ is the primary coil radius, $\rm I_p(t)$ is the excitation current, and~$\mu_0$ is the vacuum permeability. For~a sinusoidal current, ${\rm I_p(t)} = {\rm I_0}\sin(\omega {\rm t})$ with amplitude $\rm I_0$ and angular frequency $\omega$, the~resulting magnetic flux $\phi$ oscillates at the same frequency.
According to Equation~\eqref{eq:mag-field-gen}, increasing either the number of turns $\rm n_p$ or~the excitation current $\rm I_0$ strengthens the magnetic field and flux, thereby enhancing the induced differential voltage in the secondary coils. Additionally, enlarging the radius of the primary coil enhances the effective flux linkage with the secondary coils, leading to a higher induced differential voltage. However, these enhancements come at the cost of increased heat dissipation due to an increase in input voltage requirements (The voltage increases with the impedance of the coil, which is directly proportional to the number of turns and the radius.).

\begin{figure}[H]\vspace{-13pt}
\subfloat[\centering]{\includegraphics[width=0.49\linewidth]{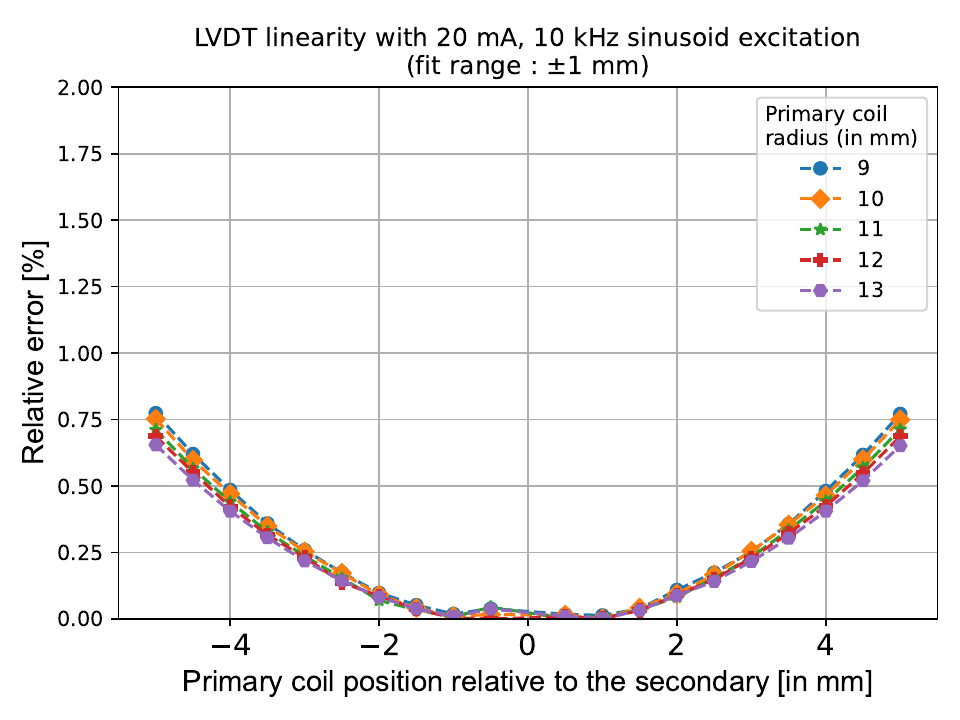}}
\subfloat[\centering]{\includegraphics[width=0.49\linewidth]{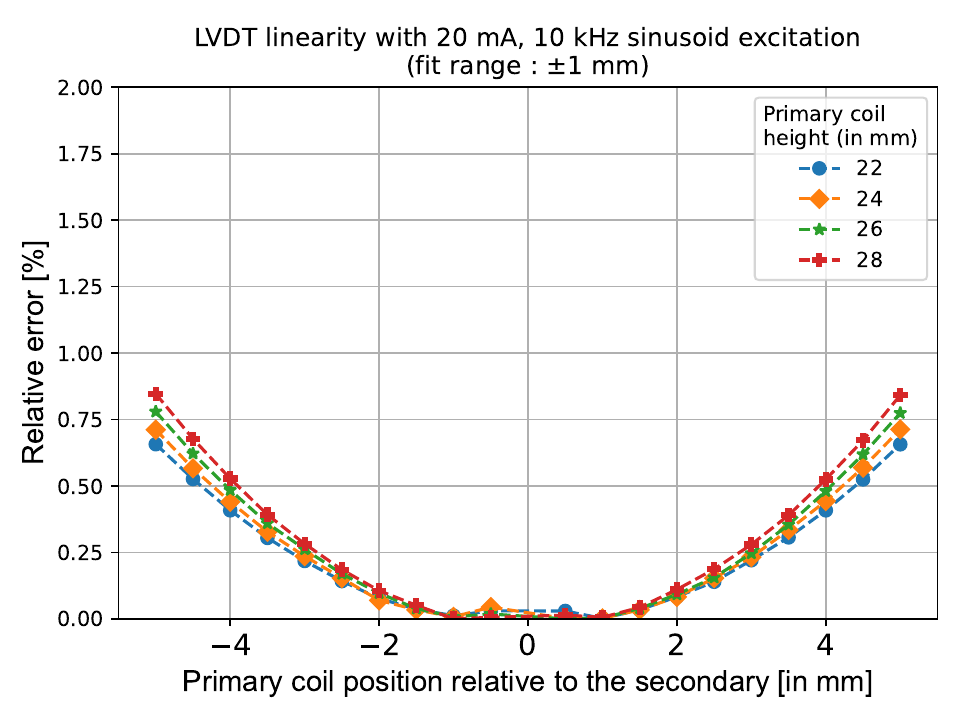}}
\caption{Influence of the primary coil dimensions on LVDT linearity when the response is fitted within a $\pm 1$~mm range. (\textbf{a}) Effect of $\rm R_p$. (\textbf{b}) Effect of $\rm H_p$. The~linearity remains largely unaffected in both~cases.}
\label{fig:inn dim lin}
\end{figure}

Interestingly, despite these variations, the~linearity of the LVDT remains largely unaffected by changes in primary coil height and radius, as~shown in Figure~\ref{fig:inn dim lin}. This is because the linearity primarily depends on the geometric symmetry and relative coil distances rather than absolute coil~dimensions.

\subsection{Magnet~Dimensions}
The size of the permanent magnet is critical to the performance of the VC actuation. A~larger magnet generates more intense magnetic fields, directly increasing the actuation force. It is therefore beneficial to maximise the magnet dimensions within the spatial constraints of the assembly. Importantly, while the magnet dimensions significantly influence the actuation force, their influence on the LVDT sensing characteristics is negligible, particularly when the dimensional variations are small relative to the original~values.

\begin{figure}[H]\vspace{-12pt}
\subfloat[\centering]{\includegraphics[width=0.49\linewidth]{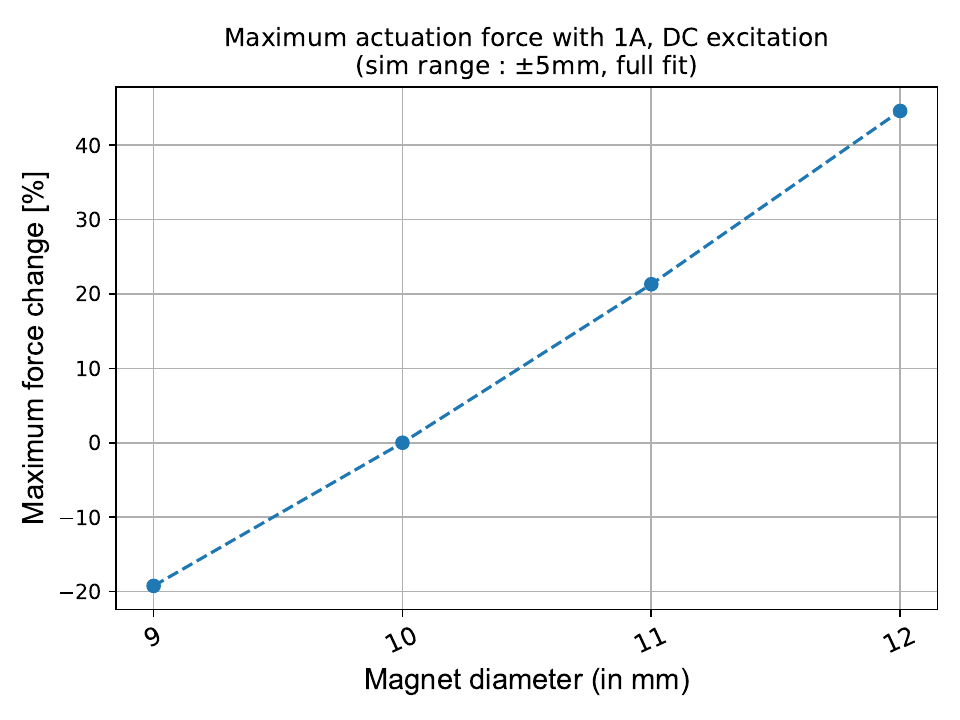}}
\subfloat[\centering]{\includegraphics[width=0.49\linewidth]{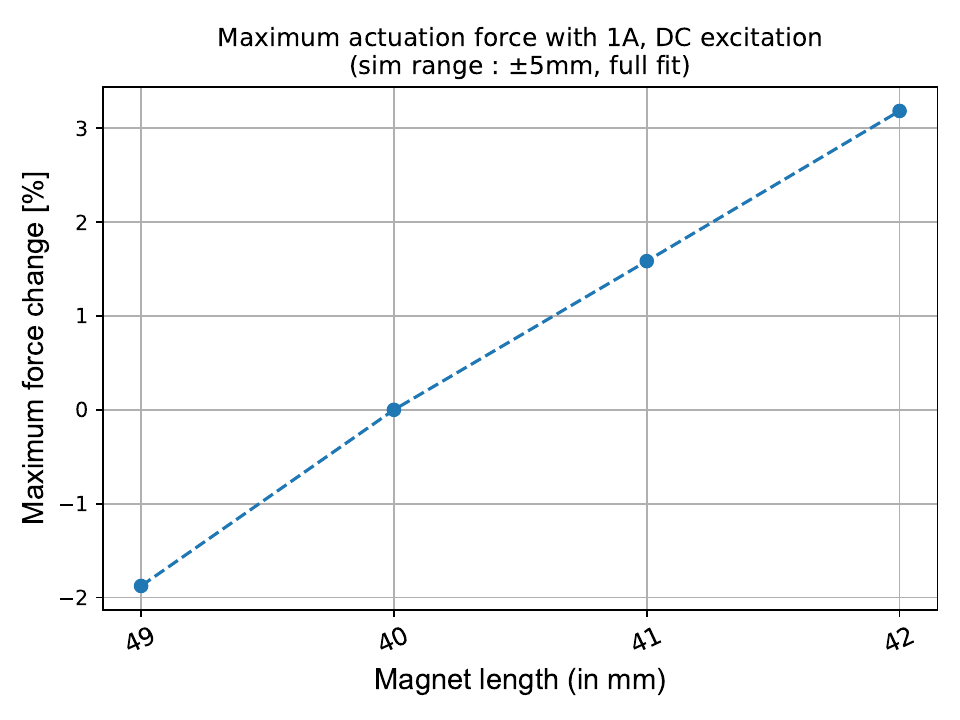}}
\caption{Effect of the magnet dimensions on VC actuation force when all other primary and secondary coil parameters remain constant. (\textbf{a}) Variation of the maximum force (in \%) with magnet diameter ($\rm 2R_m$). (\textbf{b}) Variation of the maximum force (in \%) with magnet length ($\rm H_m$). The~diameter has a much stronger influence than~length.}
\label{fig:vc response magdim}
\end{figure}

This decoupling of LVDT sensing and VC actuation mechanisms enables an independent optimisation of each function. The~results in Figure~\ref{fig:vc response magdim} clearly demonstrate that the maximum generated VC force strongly depends on the magnet size. A~millimetre increase in diameter significantly increases the force, more than 20\%, while the same increase in length improves the force by approximately 1.5\%. 

\subsection{Coil Wire~Design}
This step allows a further performance enhancement even after the geometric parameters, such as coil size and magnet dimensions, have been fixed and have reached their physical limits. Thus, there is a margin of improvement that can be achieved beyond the geometrical constraints. The~optimisation now focuses on the number of coil wire layers ($\rm N_p$, $\rm N_s$) and wire diameter ($\rm d_p$, $\rm d_s$) in each coil. Increasing the number of layers in both the primary and secondary coils improves the overall LVDT response due to the increased magnetic flux generation and reception. The~total magnetic flux $\phi$ through the secondary coil is directly proportional to the number of turns $\rm n_s$ and the magnetic field strength $\rm B$ generated by the primary coil (Equation~\eqref{eq:mag-field-gen}) and~can be expressed as follows~\cite{Griffiths:1492149}:
\begin{equation}
    \phi = \rm n_s\int_{s}{\mathbf{B}(t,z)\cdot d\mathbf{a}},
\end{equation}
where $\rm d\mathbf{a}$ is the differential area vector normal to the surface $S$ of the coil. Thus, increasing the number of turns improves the total absorbed flux $\phi$, improving the LVDT response. However, increasing the number of wire layers also leads to a higher resistance and therefore higher heat dissipation through the Joule effect. These thermal changes must be carefully controlled to maintain safe operating conditions. The~custom simulation pipeline was therefore used to evaluate the impact of additional layers on sensor performance and power dissipation. If~the number of coil wire layers is restricted by design or fabrication limits, a~comparable improvement can be achieved by reducing the wire diameter, thereby increasing the number of turns within the same coil~volume.

\begin{figure}[H]\vspace{-10pt}
\subfloat[\centering]{\includegraphics[width=0.49\linewidth]{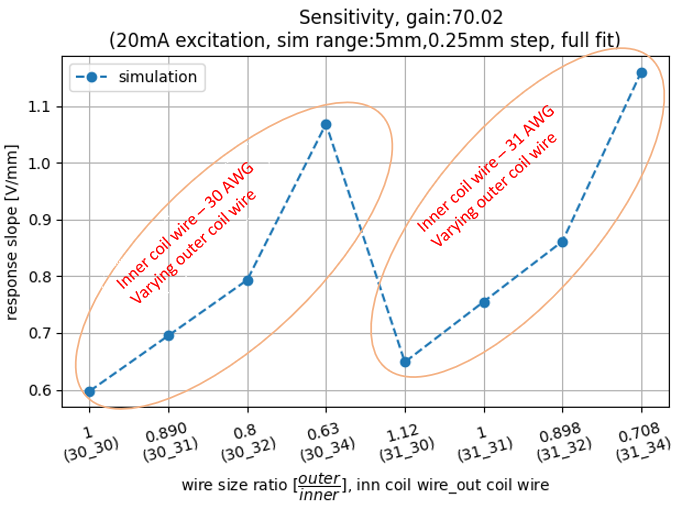}}
\subfloat[\centering]{\includegraphics[width=0.49\linewidth]{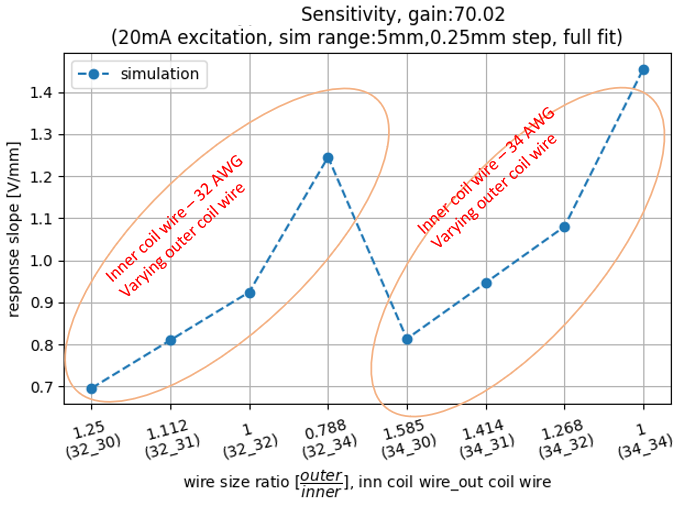}}
\caption{Effect of varying primary and secondary coil wire diameter ($\rm d_p, d_s$) on the non-normalised LVDT response. The~x-axis labels describe the ratio of $\rm d_s$ over $\rm d_p$, followed by the wire-gauge combinations: i.e.,~(30\_31) indicates a $\rm d_p = 30$ AWG primary and $\rm d_s = 31$ AWG secondary coil wire diameter. (\textbf{a}) Simulation results with $\rm d_p = 30, 31$ AWG; (\textbf{b}) simulation results with \mbox{$\rm d_p = 32, 34$ AWG.} A~thinner wire enables more turns for fixed coil heights, which increases the LVDT response. The~secondary coil wire diameter has a larger impact than the primary coil wire~diameter.}
\label{fig:wire thickness}
\end{figure}

The simulation results summarised in Figure~\ref{fig:wire thickness} show that reducing the wire diameter enhances the non-normalised LVDT response, an~effect that is more pronounced for the secondary coil than for the primary coil. When $\rm d_s$ is decreased from 30 AWG to 34 AWG, while $\rm d_p$ remains constant, an~improvement of 78\% in response is observed. Decreasing $\rm d_p$ from 30 AWG to 34 AWG, with~$\rm d_s$ constant, results in a 33\% improvement. For~fixed coil heights ($\rm H_p, H_s$), using thinner wire allows a larger number of turns ($\rm n_p, n_s$), which increases the total magnetic flux linkage and hence the output~signal.

\begin{figure}[H]
\subfloat[\centering]{\includegraphics[width=0.49\linewidth]{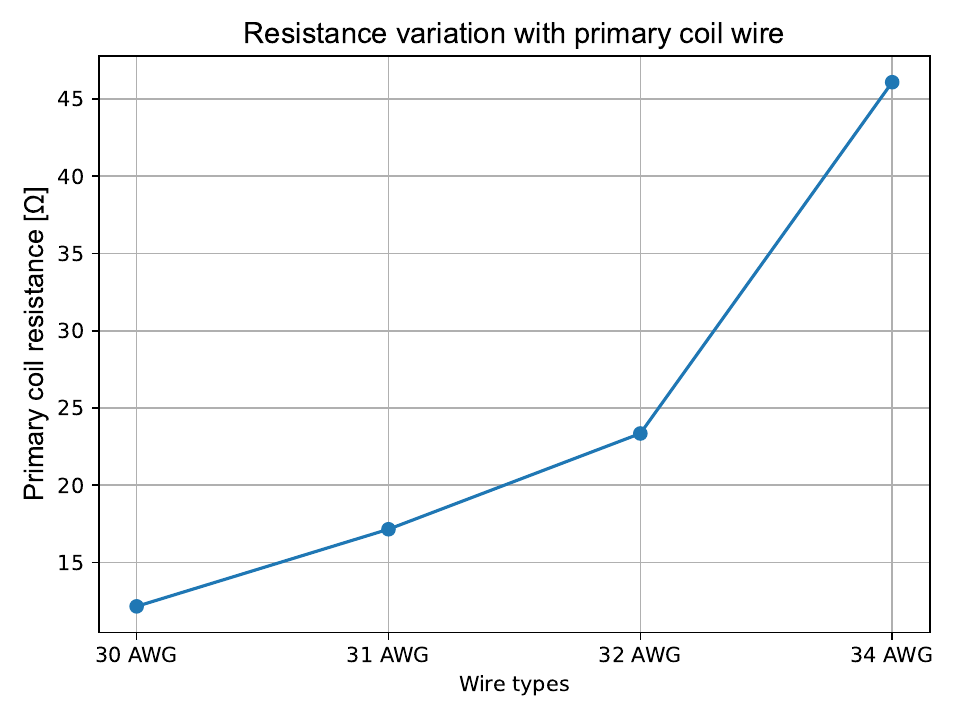}}
\subfloat[\centering]{\includegraphics[width=0.49\linewidth]{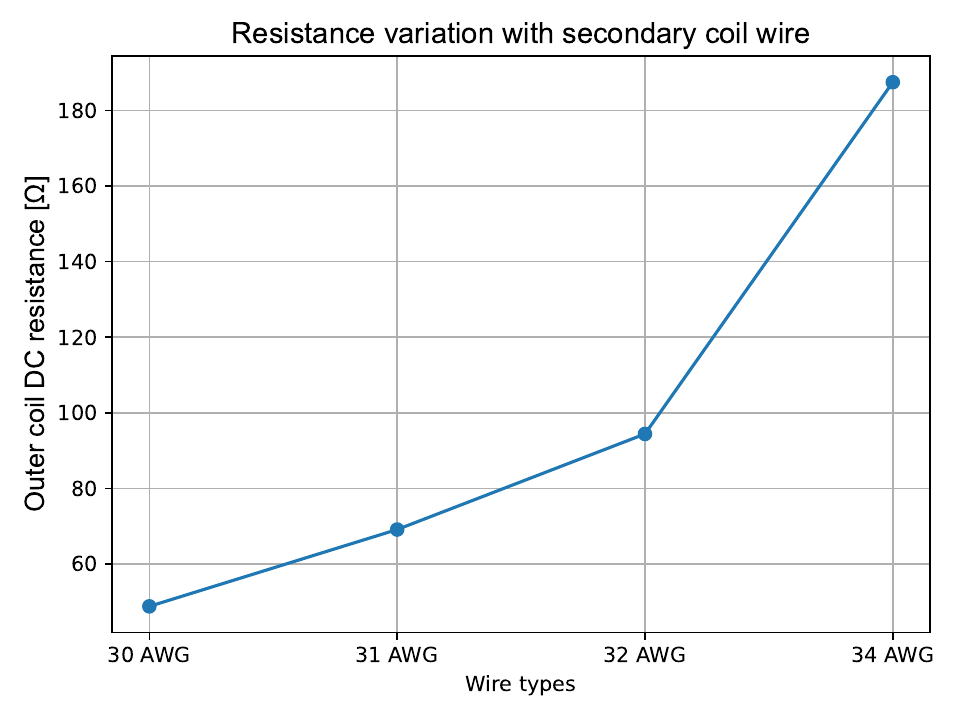}}
\caption{Resistance variation with coil wire diameter. (\textbf{a}) Primary coil; (\textbf{b}) secondary coil. The~resistance increases with smaller diameters due to the $R \propto 1/\rm d^2_{p,s}$ dependence and longer total length per layer. Managing resistance is crucial for balancing response improvements against thermal and electronic~constraints.}
\label{fig:wire impedance}
\end{figure}

However, reducing the wire diameter also increases the DC resistance (and thus impedance) of the coil, as~shown in Figure~\ref{fig:wire impedance}, due to its dependence on the wire cross-section area and total length. This raises the primary coil voltage and can lead to additional heat generation. It also affects the normalised LVDT response, in~particular when the primary coil is driven by a voltage source. Therefore, the~voltage-normalised response, expressed as $\rm [V_{out}/mmV_{in}]$, provides a more meaningful measure of overall~performance.

\vspace{-21pt}

\begin{figure}[H]\vspace{-12pt}
\subfloat[\centering]{\includegraphics[width=0.49\linewidth]{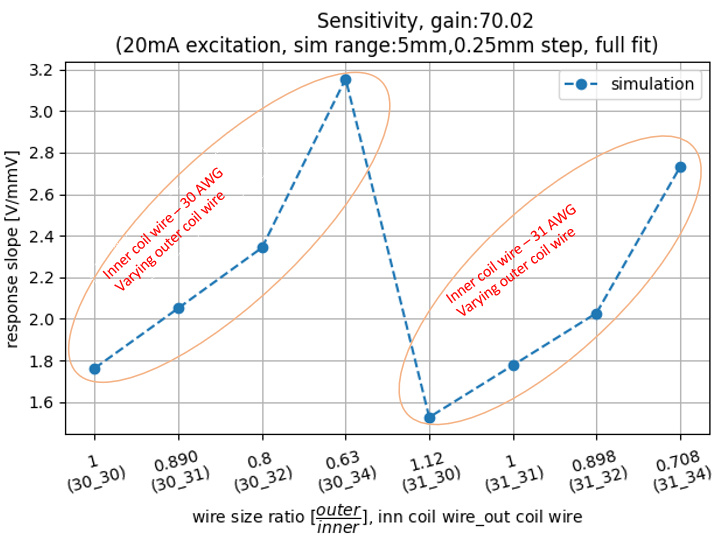}}
\subfloat[\centering]{\includegraphics[width=0.49\linewidth]{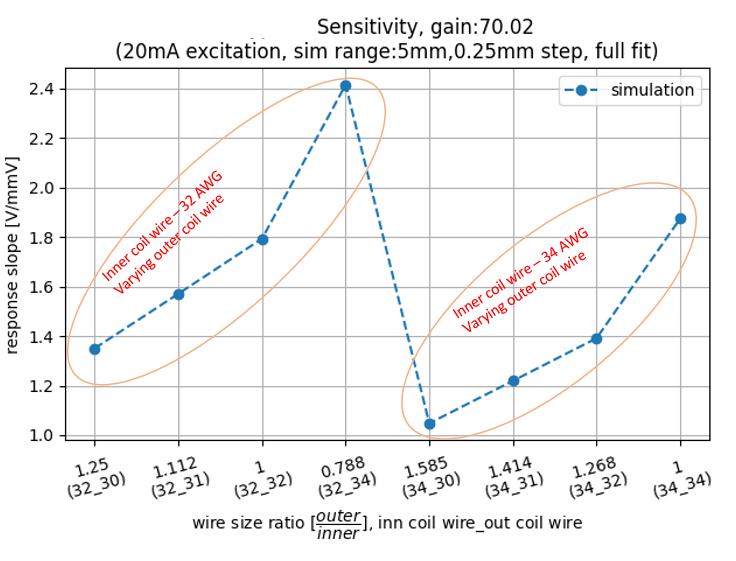}}
\caption{Effect of varying primary and secondary coil wire diameter ($\rm d_p, d_s$) on the voltage-normalised LVDT response. The~x-axis labels describe the ratio of $\rm d_s$ over $\rm d_p$, followed by the wire-gauge combinations: i.e.,~(30\_31) indicates a $\rm d_p = 30$ AWG primary and $\rm d_s = 31$ AWG secondary coil wire diameter. (\textbf{a}) Simulation results with $\rm d_p = 30, 31$ AWG; (\textbf{b}) simulation results with \mbox{$\rm d_p = 32, 34$ AWG.} A~decrease of $\rm d_p$ now results in a lower LVDT~response.}
\label{fig:wire response}
\end{figure}

This is presented in Figure~\ref{fig:wire response}, which shows that when $\rm d_s$ is decreased from 30 AWG to 34 AWG (with $\rm d_p$ constant), an improvement of 79\% in voltage-normalised response is observed, similar to the results in Figure~\ref{fig:wire thickness}. However, decreasing $\rm d_p$ from 30 AWG to 34 AWG (with $\rm d_s$ constant) now results in a 40\% lower voltage-normalised LVDT response. This is because the increase in resistance in the primary coil outweighs the LVDT response improvement, resulting in a reduced net response per input voltage. Balancing the acceptable coil impedance against the normalised LVDT response is thus crucial for selecting the appropriate wire type. Additionally, mechanical considerations such as wire tensile strength and production tolerances play a significant role in this selection. The~developed simulation pipeline supports this selection by predicting the electromagnetic and thermal outcomes for different wire types. The~LVDT linearity remains largely unaffected by variations in wire size, with~<$1\%$ changes, as determined by the geometric coil~parameters. 

The number of charge carriers flowing through the secondary coils increases with the number of turns for a constant DC current. This means that more turns, i.e.,~smaller wire diameter $\rm d_s$ or more layers $\rm N_s$, will directly enhance the actuation force. The~total force $\rm \mathbf{F}$ generated on these charge carriers (or wire) due to the magnetic field can be expressed as~\cite{Jackson:1998nia}:
\begin{equation}
    \rm \mathbf{F} = n_s I_s\int_{L}(\mathbf{B} \times d\mathbf{l}),
\end{equation}
where $\rm I_s$ represents the current through the secondary coils, $\rm L$ is the length of the coil wire of one turn, $\rm d\mathbf{l}$ is the differential length vector along the coil, and~$\rm n_s$ is the total number of turns. As~the number of turns increases, the~force $\rm \mathbf{F}$ increases accordingly due to the higher interaction between the magnetic field and the current-carrying~coils.
\section{Experimental validation}
\label{sec:experimental_validation}

To validate the new methodology, presented in the previous sections, it is applied to a specific use case in ETpathfinder. Starting from an available initial design (summarised in Table~\ref{tab:parameter values}) of a combined LVDT+VC system, developed for the small inverted pendulum stage of the ETpathfinder mirror tower seismic isolation~\cite{ETpathfinderTDR}, the~optimisation sequence described in Section \ref{sec:methodology} is used to further improve the performance of the sensor and actuator. The~mechanical constraints are set to $\rm R_{env} = 20$~mm and $\rm H_{env} = 60$~mm, while within the dynamic range ($\pm 2.5$~mm), the requirements on LVDT linearity and VC force stability are set to $\geq$99\% and $\geq$97\%, respectively. A~prototype is then constructed with the optimised design parameters and~fully characterised with a dedicated experimental setup~\cite{setup-paper} at the University of Antwerp. This allows us to validate the simulation results of the optimised design with actual prototype measurements and~provides an indication how manufacturing tolerances, material inconsistencies, or~thermal variations could influence the performance by evaluating the statistical and systematic~uncertainties.

\begin{figure}[H]\vspace{-5pt}
\subfloat[\centering]{\includegraphics[width=0.49\linewidth]{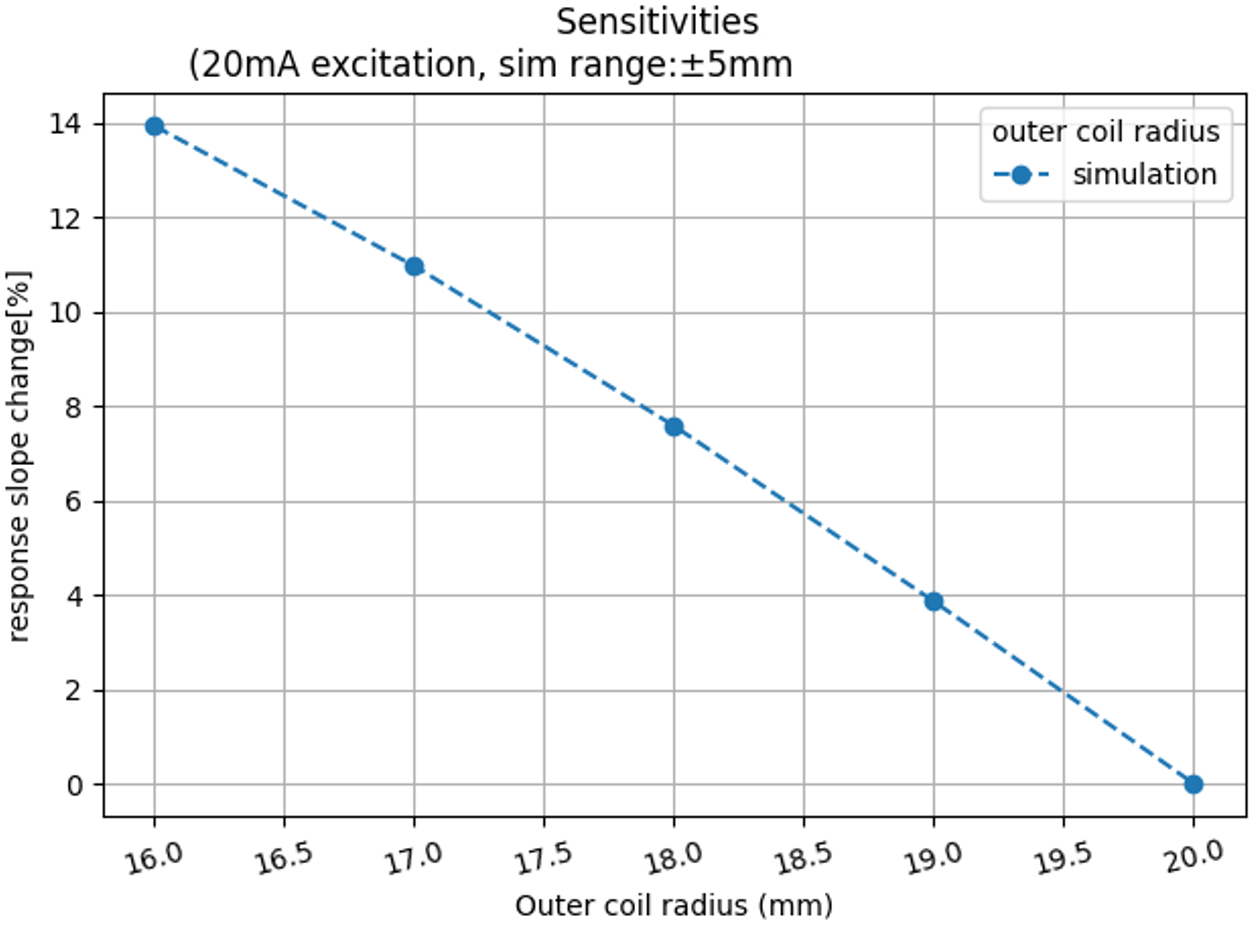}}
\subfloat[\centering]{\includegraphics[width=0.49\linewidth]{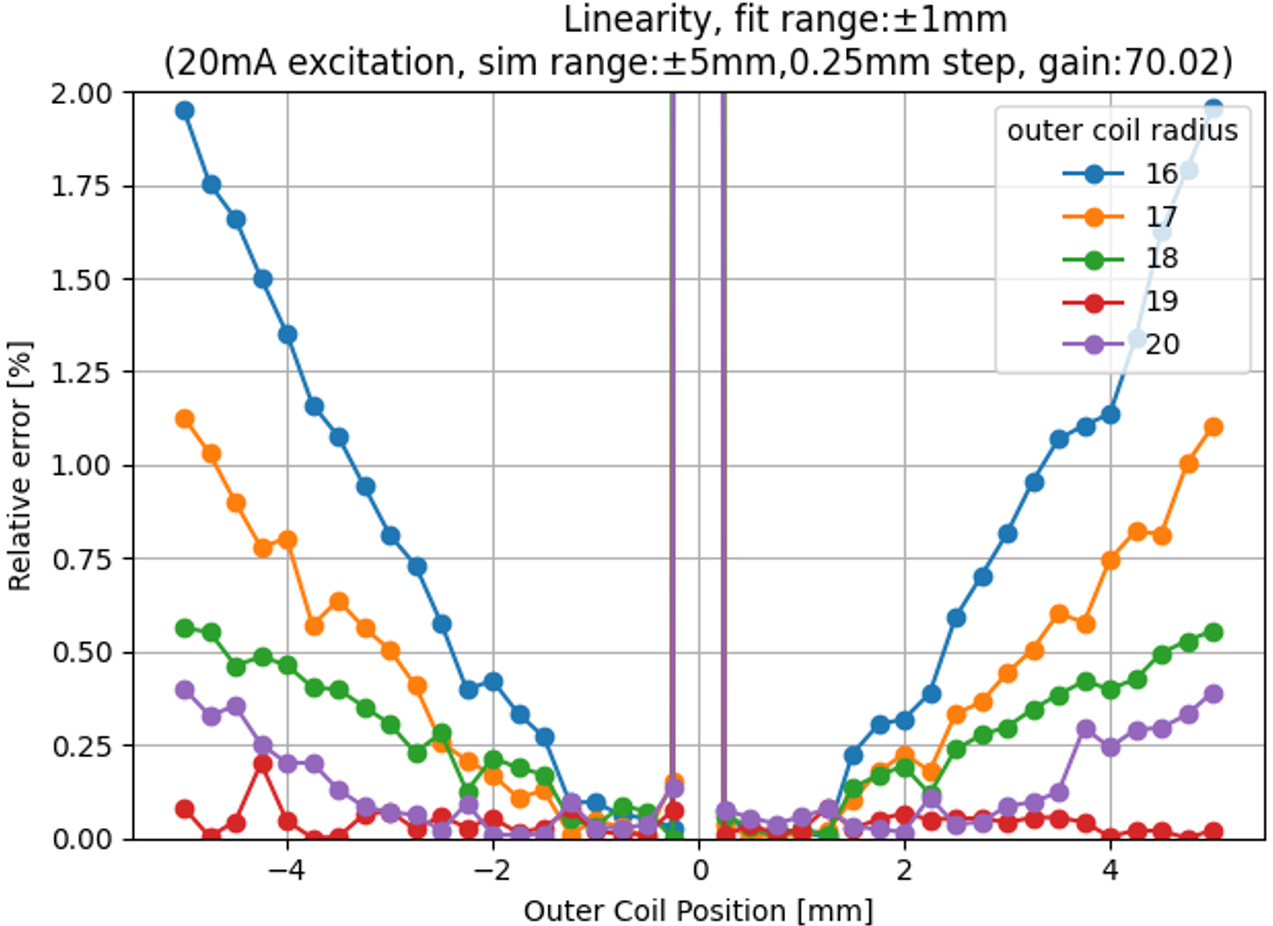}}
\caption{Effect 
 of $\rm R_s$ on LVDT response (\textbf{a}) and linearity (\textbf{b}) when fitted within a $\pm 1$~mm range. The~initial design ($\rm R_s = 20$~mm) is improved by reducing $\rm R_s$ to 17~mm, which enhances LVDT response while maintaining linearity above the 99\% threshold. A~further reduction to 16 mm yields a higher response but degrades linearity significantly. This example illustrates the methodology used for all parameters, where improvements are balanced against operational and mechanical constraints specific to ETpathfinder. The~fluctuations visible in linearity are due to the applied voltage normalisation of the LVDT~response.}
\label{fig:topup lvdt improvements}
\end{figure}

To illustrate the procedure, the~optimisation of the secondary coil radius $\rm R_s$ is presented, as it clearly influences both VC actuation force and LVDT response. The~simulation results in Figure~\ref{fig:topup lvdt improvements} show the impact of $\rm R_s$ on the LVDT response. As~expected, reducing the radius increases the response due to stronger magnetic coupling. However, this improvement comes at the cost of degraded~linearity.

Reducing $\rm R_s$ from 20~mm (initial design) to 17~mm yields a significant improvement in response while maintaining linearity above the 99\% threshold. A~further reduction to 16~mm yields another 3\% improvement in response, but the linearity drops below the acceptable limit, and~the available transverse motion ($\Delta \rm R_{space}$) is further reduced. Based on this trade-off, the~final optimised design adopts $\rm R_s =$ 17~mm, giving the best balance between enhanced LVDT response and linear performance. This demonstrates the central principle of the optimisation procedure: parameters are not optimised in isolation but are chosen considering both operational constraints and performance standards. For~GW detectors, these include strict requirements on linearity, response, heat dissipation, and~spatial compatibility with the suspension system. While $\rm R_s$ is shown as a representative case, the~same approach is applied to all design parameters. The~optimised configuration is then obtained by systematically balancing performance improvements with the practical constraints and requirements set by ETpathfinder. The~optimised design’s final parameters are summarised in Table~\ref{tab:parameter values} and yield a 2.8-fold improvement in LVDT response and a 2.5-fold increase in the maximum VC force compared to the initial~design.

\begin{table}[H] 
\caption{Summary of initial and optimised design parameters for the combined LVDT+VC system to be used in the mirror seismic isolation of ETpathfinder. The~optimised values were obtained following the methodology introduced in Section~\ref{sec:methodology} using the mechanical constraints and performance requirements set by~ETpathfinder.}
\label{tab:parameter values}
\begin{tabularx}{\textwidth}{>{\hsize=1.2\hsize}C>{\hsize=0.9\hsize}C>{\hsize=0.9\hsize}C}
\toprule
\textbf{Parameter}	& \textbf{Initial Dimensions}  & \textbf{Optimised Dimensions}\\
\midrule
Secondary coil distance ($\rm D_s$) & 40.0~mm & 35.0~mm \\ 
Secondary coil radius ($\rm R_s$) & 20.0~mm & 17.0~mm \\ 
Secondary coil height ($\rm H_s$) & 10.0~mm & 25.0~mm \\ 
Secondary coil layers ($\rm N_s$) & 5 & 5 \\ 
Primary coil radius ($\rm R_p$) & 9.0~mm & 7.0~mm \\ 
Primary coil height ($\rm H_p$) & 20.0~mm & 20.0~mm \\ 
Primary coil layers ($\rm N_p$) & 6 & 6 \\ 
Magnet radius ($\rm R_m$) & 4.0~mm & 4.0~mm \\ 
Magnet height ($\rm H_m$) & 30.0~mm & 30.0~mm \\ 
Magnet type & NdFeB N42 & NdFeB~N42 \\ 
Coil wire diameter ($\rm d_p, d_s$) & 32 AWG & 32~AWG \\ 
\bottomrule
\end{tabularx}
\end{table}

\subsection{LVDT~Characterisation}
Following the measurement procedure described in \cite{setup-paper}, LVDT response measurements were performed by displacing the primary coil along its axis symmetrically about the approximate electrical zero, covering a range of $\pm2.5$~mm, and~reading out the induced differential voltage on the secondary coils with an oscilloscope.
This is achieved by fixing the primary coil to a vertical linear translation stage, while the secondary coil assembly is suspended on a custom mount that enables a precise horizontal and vertical alignment of the coils. The~primary coil is driven with a voltage source and~excited with a sinusoidal signal of frequency 10~kHz and amplitude 2.5~V. The~measured data were fitted with a linear polynomial independently for the positive and negative halves of the motion, and~the mean slope parameter was taken as the effective response. These results are then divided by the primary coil input voltage to obtain the normalised LVDT response value. The~measurement is repeated three times to obtain a statistical uncertainty. In~addition, the~considered systematic uncertainties arise from the following four~sources: 
\begin{itemize}
    \item \textbf{Excitation voltage
}: directly scales the sensor response, and~as the choice of the actual value depends on the DAQ system and sensor application, we modified the default excitation value (2.5~V) to 2.25~V and 2.9~V to study the stability of the response after voltage normalisation.
    \item \textbf{Transverse offset}: reflects potential transverse coil misalignments in the experimental measurement setup. Therefore, the primary coil transverse position was modified with $\pm 0.5$~mm.  
    \item \textbf{Step size}: refers to the measured points of primary coil displacement and~influences the linear fit results. To~study how the sensor response is affected by this choice, we modified the default step size (0.5~mm) to 0.25~mm and 1.0~mm.
    \item \textbf{Mesh resolution}: determines the accuracy of the field gradients in the simulation model. The~default setting (0.5) of Airspace-2 was adjusted to 0.1 (very fine grid) and 1.0 (coarse grid) to study the influence on LVDT response.
\end{itemize}

\begin{figure}[H]\vspace{-14pt}
\subfloat[\centering]{\includegraphics[width=0.49\linewidth]{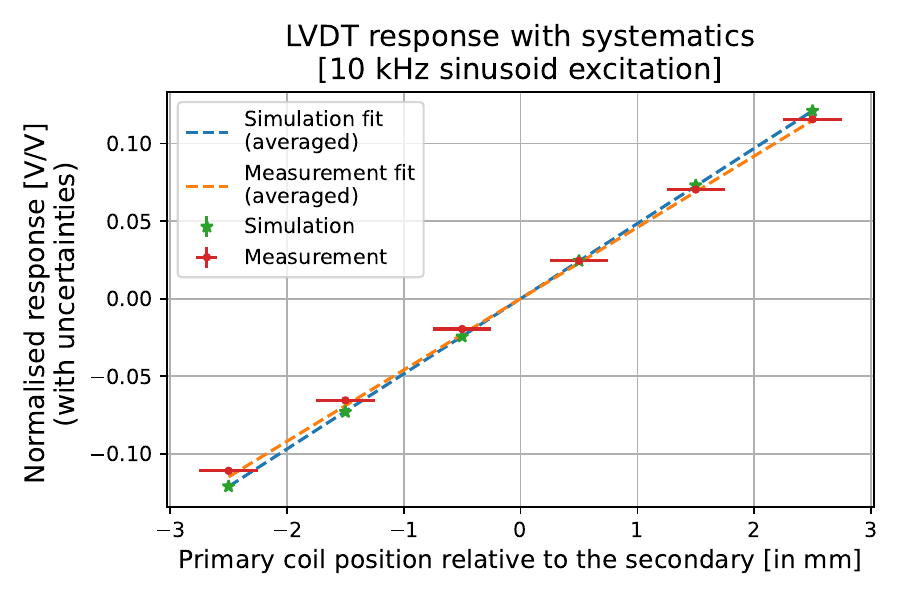}}
\subfloat[\centering]{\includegraphics[width=0.49\linewidth]{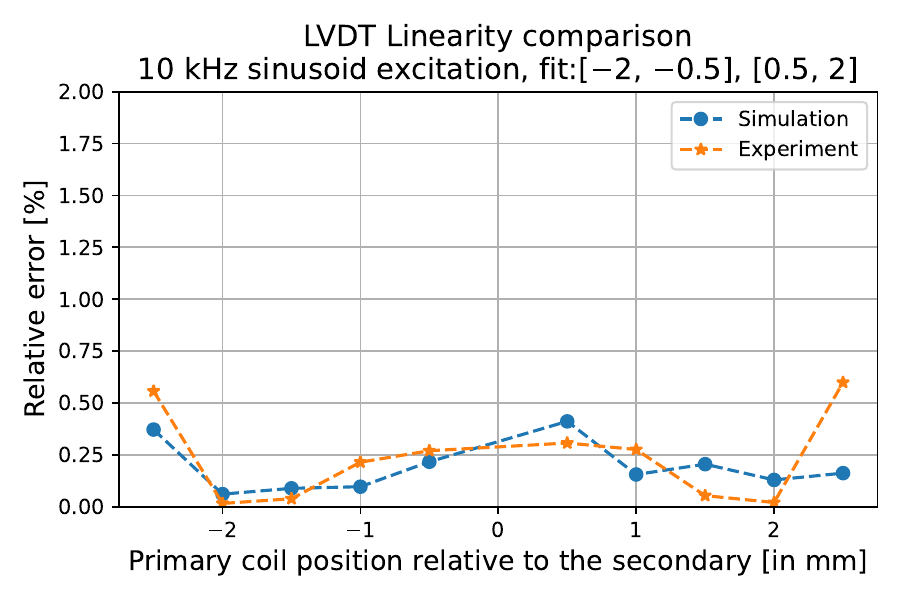}}
\caption{Voltage-normalised response (\textbf{a}) and linearity (\textbf{b}) of the optimised LVDT when the primary coil is excited with a 2.5~V 10~kHz sinusoid. Independent linear fits over a 1.5~mm range on the negative and positive halves of the motion were averaged to account for noise obscuring the true centre. The~measured response shows a good agreement with the simulation, with~overall linearity exceeding 99\%.}
\label{fig:topup response comparison}
\end{figure}

The excitation voltage and transverse offset systematics are only applied to data, while the mesh resolution systematic only applies to simulations. In the present work, only the resulting combined uncertainties are summarised. A detailed explanation of these effects is provided in \cite{setup-paper,kumar-thesis}. The measured response and linearity are shown in Figure~\ref{fig:topup response comparison} and compared with a simulation of the optimised design. The experimentally measured value is found to be $0.0460^{+0.0006}_{-0.0009}$~V/mmV, while the simulated response is found to be $0.0485^{+0.0006}_{-0.0009}$~V/mmV. The~measured response has a total uncertainty of +1.3\% and $-$1.9\%, dominated by the excitation voltage systematic source, while the statistical uncertainty is negligible (0.05\%). The~simulated response has a total uncertainty of +1.2\% and $-$1.8\%, dominated by the mesh resolution systematic source. This agreement, within~5\% of the predicted value, demonstrates the predictive accuracy of the simulation framework and confirms that the optimised prototype performs in line with the modelled~expectations.

\subsection{VC~Characterisation}
The Voice Coil actuator force measurements were carried out over the same displacement range of $\pm2.5$~mm, ensuring consistency with the LVDT characterisation, and~following the same procedure introduced in \cite{setup-paper}. The~secondary coils are driven with a low-frequency ($0.1$~Hz) sine wave with an amplitude of $30.75$~mA, and~at each position of the primary coil, the apparent weight of the suspended secondary coil assembly is measured by a precision balance placed below the custom mount. This weight change follows the 0.1~Hz signal due to the produced Lorentz force and~allows us to calculate the normalised force (N/A) through the gravitational acceleration ($\rm \mathbf{g} = 9.81\ m/s^2$). To~tune the weight on the precision balance, the secondary coil assembly is supported by three springs. This introduces a correction factor $k = 1.171 \pm 0.016$~\cite{setup-paper} to take the mechanical coupling into account. The~final VC actuator force is then calculated via $\mathbf{F} = k\Delta m\mathbf{g}$, where~$\Delta m$ is the amplitude of the weight changes measured by the precision balance. The~measurement is repeated three times to obtain a statistical uncertainty, and~the following systematic uncertainties are~considered:  
\begin{itemize}
    \item \textbf{Excitation voltage}: directly scales the produced force, and~as the choice of the actual value depends on the DAQ system and actuator application, we modified the default voltage amplitude (5~V) that drives the current source by $\pm 1$~V to study how it affects the normalised force.
    \item \textbf{Transverse offset}: reflects potential transverse coil misalignments in the experimental measurement setup. Therefore, the primary coil transverse position was modified with $\pm 0.5$~mm.  
    \item \textbf{Spring correction factor}: to propagate the uncertainty of the correction factor $k$ to the calculated actuator force, we vary this factor with its error $\pm 0.016$.
    \item \textbf{Mesh resolution}: determines the accuracy of the field gradients in the simulation model. The~default setting (0.5) of Airspace-2 was adjusted to 0.1 (very fine grid) and 1.0 (coarse grid).
\end{itemize}

The excitation voltage, transverse offset, and~spring correction factor systematics are applied to data, while the mesh resolution systematic only applies to~simulations.

\begin{figure}[H]
\subfloat[\centering]{\includegraphics[width=0.49\linewidth]{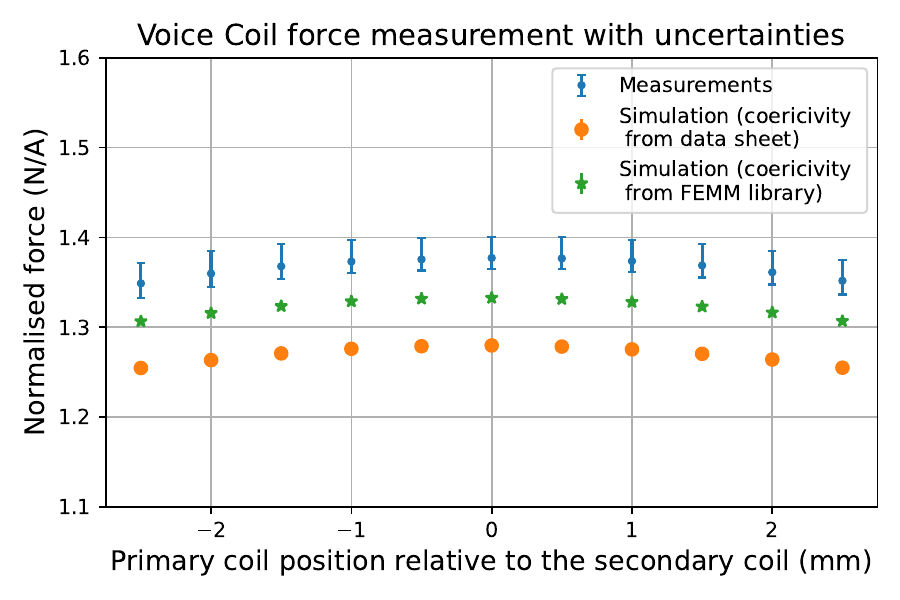}}
\subfloat[\centering]{\includegraphics[width=0.49\linewidth]{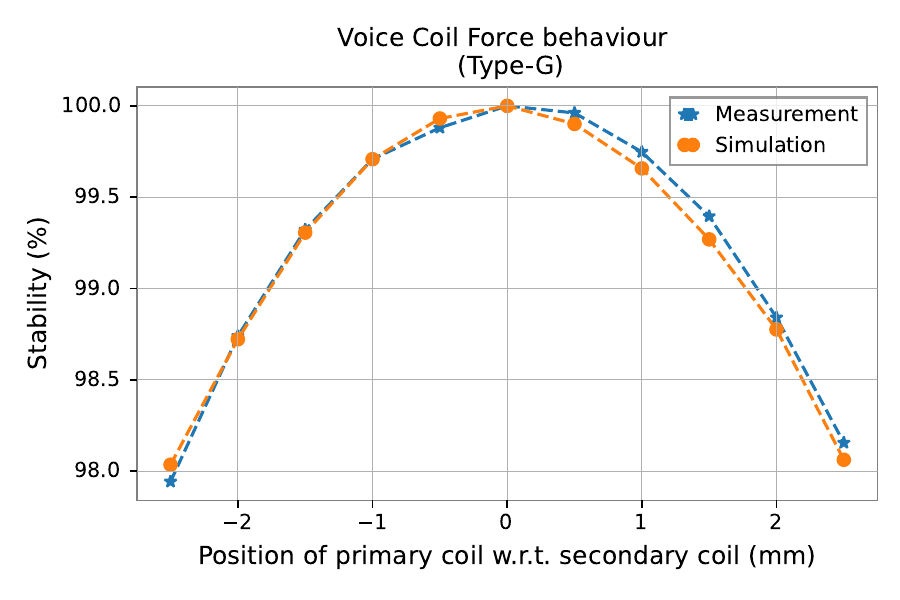}}
\caption{Comparison of simulation and measurement results for the VC actuator: (\textbf{a}) normalised force, (\textbf{b}) force stability across the measured $\pm2.5$ mm range. The~measured force is in close agreement with the simulations, which clearly depend on the used magnet coercivity value. Both simulation and measurement show more than 98\% stability across the dynamic range, and~more than 99\% within $\pm 1$~mm, confirming a consistent actuator~performance.}
\label{fig:topup vc force}
\end{figure}

The measured VC force and VC force stability of the prototype are shown in Figure~\ref{fig:topup vc force} and~compared with simulations of the optimised design. The~blue, orange, and~green datasets correspond respectively to the experimentally measured forces (including all systematic uncertainties), the~simulated forces with manufacturer‑specified magnet coercivity (955,000~A/m), and~the simulated forces with the default magnet coercivity of the FEMM library (994,529~A/m). The~dominant source of uncertainty arises from the spring correction factor $k$. The~experimental data reach a maximum force of $1.3771^{+0.0236}_{-0.0119}~\mathrm{N/A}$, while simulations with the default coercivity reach $1.3325^{+0.0002}_{-0.0004}~\mathrm{N/A}$, and~simulations with a coercivity specified by the magnet manufacturer reach $1.2796^{+0.0002}_{-0.0004}~\mathrm{N/A}$. The~maximum measured force has a total uncertainty of +1.72\% and $-$0.87\%, whereas the maximum simulated forces exhibit a negligible total uncertainty of +0.02\% and $-$0.03\%. The~measurement exceeds the simulation by 7.1\% (manufacturer data sheet coercivity) and 3.4\% (default FEMM library coercivity), respectively. This discrepancy cannot be attributed solely to systematic uncertainties and instead highlights limitations in the material models. In~particular, the~magnetic properties of the permanent magnet such as coercivity, residual magnetisation, and~possible inhomogeneities may deviate from the nominal values specified by the manufacturer. The~close correspondence between the default FEMM coercivity simulation and experiment confirms the effectiveness of the optimisation~procedure.

\vspace{-15pt}

A comparison of key performance metrics for the initial and optimised configurations is summarised in Table~\ref{tab:optimised values}. The~optimised design exhibits major improvements in LVDT response and VC actuation force while maintaining excellent linearity and stability. These results confirm that the simulation-based optimisation framework presented in Sections~\ref{sec:simulation_framework} and \ref{sec:methodology} reliably predicts both sensing and actuation performance and~offers a validated methodology for designing high-precision LVDT+VC systems suitable for advanced applications such as GW~detectors.

\begin{table}[H] 
\small
\caption{Comparison of simulated and measured results for the initial and optimised designs of the LVDT+VC system for ETpathfinder. The~optimised design shows significant improvements in LVDT response ($\times$2.8) and VC force ($\times$2.5) while maintaining linearity and force stability, with~a good agreement between simulations and prototype~measurements.}
\label{tab:optimised values}
\begin{tabularx}{\textwidth}{>{\hsize=1.4\hsize}L>{\hsize=0.7\hsize}C>{\hsize=0.95\hsize}C>{\hsize=0.95\hsize}C}
\toprule
\textbf{Parameter}	& \textbf{Initial Design (Simulation)}  & \textbf{Optimised Design (Simulation)}  & \textbf{Optimised Design (Measurement)}\\
\midrule
LVDT Response $\left(\rm \frac{V_{out}}{mm\cdot V_{in}}\right)$ & 0.017  & 0.048  & 0.046 \\
LVDT Linearity (\%) & >$99$  & >$99$  & >$99$ \\
VC Force $\left(\rm \frac{N}{A}\right)$ & 0.523  & 1.333  & 1.377 \\
VC Force Stability (\%) & >$98$  & >$98$  & >$98$ \\ 
\bottomrule
\end{tabularx}
\end{table}
\section{Conclusions}
\label{sec:conclusions}

This paper introduces a novel optimisation methodology to study and design combined LVDT and VC systems. It uses a custom-developed software pipeline that is based on finite element simulations to implement the coil geometries and systematically analyse the performance of the proposed design. We demonstrated the optimisation sequence with simulation studies and~applied it to a specific use case in gravitational wave instrumentation. With our proposed methodology, we improved a combined LVDT+VC system for the ETpathfinder seismic isolation system. A~prototype of the optimised design was constructed and experimentally validated with dedicated measurements. It reached a 2.8-fold increase in LVDT response and a 2.5-fold increase in VC actuation force with respect to the initial design, while LVDT linearity and VC force stability were~maintained.

The proposed methodology achieves balanced improvements in LVDT response, linearity, VC actuation force, and~stability while respecting practical thermal and geometric constraints. By~explicitly revealing key parameter dependencies, such as the competing effects of secondary coil distance on LVDT linearity and response, and~the influence of the radial air gap on sensing and actuation, the~simulation framework enables one to obtain a complete picture of the system behaviour and to identify application-specific design~trade-offs.

A central contribution of this work lies in the integration of sensing and actuation optimisation within a single, unified workflow. Unlike existing approaches that treat LVDT sensors and VC actuators as separate design problems, the~proposed framework allows the two subsystems to be optimised independently or jointly without requiring a redesign. This flexibility enables operation as a stand-alone LVDT, a~VC actuator, or~a combined sensor--actuator system. The~optimisation process follows a structured sequence. Design begins with the secondary coils, where secondary coil distance and radius are selected to meet linearity requirements, followed by optimisation of the primary coil radius taking into account the requirement set on the radial gap. Then, the secondary coil height is tuned to maximise magnetic flux reception while maintaining mechanical feasibility. The~primary coil is further designed to maximise flux generation, taking into account power dissipation constraints. After~this, the~magnet dimensions can be maximised within the primary coil. When geometric limitations restrict further improvements, performance gains can still be achieved through adjustments in the number of coil wire layers and coil wire diameter. These steps, implemented through a custom \texttt{pyFEMM}-based finite element simulation pipeline, collectively ensure high-performance LVDT and VC~designs.

The results demonstrate the trade-offs required to simultaneously optimise LVDT response and linearity alongside VC force generation and stability. By~offering a unified parameter-driven design methodology, this work addresses a critical gap in the literature and provides a robust foundation for the development of precision sensing and actuation systems. Future research may extend this framework to include magnetic shielding effects, non-linear material behaviour, multi-objective optimisation, and~machine learning-based techniques to further enhance system performance and adaptability. Although~the focus in this paper is on the development of combined LVDT+VC systems for gravitational wave detectors, the~simulation framework and optimisation methodology can also be employed to develop similar systems for other applications in research (e.g., particle accelerators) or industry. The~same optimisation principles apply to conventional industrial LVDTs with moving ferromagnetic cores, as~the underlying electromagnetic behaviour remains unchanged. The~observed agreement between simulation and experiment further supports the general validity of the approach. Therefore, the~proposed framework can be extended to a broad range of sensing and actuation~applications.

\subsection*{Funding}{The ETpathfinder project in Maastricht is funded by Interreg Vlaanderen-Nederland, the province of Dutch Limburg, the province of Antwerp, the Flemish Government, the province of North Brabant, the Smart Hub Flemish Brabant, the Dutch Ministry of Economic Affairs, the Dutch Ministry of Education, Culture and Science, and by own funding of the involved partners.
In addition the ETpathfinder team acknowledges support from the European Research Council (ERC), the Dutch Research Council (NWO), the Research Foundation Flanders (FWO), the German Research Foundation (DFG), Spanish MICINN, the CERCA program of the Generalitat de Catalunya and the Dutch National Growth Fund (NGF).}

\subsection*{Availability}{The simulation framework developed for this study and used during the optimisation studies is available here: \url{https://github.com/beginner117/LVDT-VC-modelling-toolkit/tree/simulation_paper}, which corresponds to the version used in this paper.} 

\subsection*{Acknowledgments}{The authors would like to thank the researchers at Nikhef and the ETpathfinder collaboration, especially Fred Schimmel, Alessandro Bertolini, Bas Swinkels, Mathijs Baars, and Stefan Hild for the valuable discussions and their input to the requirements of the combined LVDT+VC system.}

\bibliographystyle{unsrt}
\bibliography{references}

\end{document}